%% file: main.tex
\documentclass[%
reprint,
superscriptaddress,
 amsmath,amssymb,
 aps,
]{revtex4-2}

\usepackage{graphicx}
\usepackage{dcolumn}
\usepackage{bm}
\usepackage{natbib}
\usepackage{multirow}
\usepackage{xcolor}
\definecolor{iqmblue}{RGB}{117,157,235}
\definecolor{iqmgreen}{RGB}{95, 221, 151}
\usepackage[colorlinks = true,
            linkcolor = iqmblue,
            urlcolor  = iqmblue,
            citecolor = iqmblue,
            anchorcolor = iqmblue]{hyperref}

\usepackage[normalem]{ulem}

\newcommand{\add}[1]{#1}
\newcommand{\rem}[1]{}

\newcommand{\cz}{CZ}

\usepackage[acronym]{glossaries}
\glsdisablehyper

\newacronym{gst}{GST}{Gate Set Tomography}
\newacronym{rb}{RB}{Randomized Benchmarking}
\newacronym{xeb}{XEB}{Cross-Entropy Benchmarking}
\newacronym{qpu}{QPU}{Quantum Processing Unit}
\newacronym{rc}{RC}{Randomized Compiling}
\newacronym{pt}{PT}{Pauli Twirling}
\newacronym{spam}{SPAM}{State Preparation and Measurement}
\newacronym{dof}{DOF}{Degrees of Freedom}
\newacronym{cb}{CB}{Cycle Benchmarking}
\newacronym{mlcb}{MLCB}{Multi-Layer Cycle Benchmarking}
\newacronym{tq}{2Q}{\add{two-qubit}}
\newacronym{sq}{1Q}{\add{single-qubit}}
\newacronym{pec}{PEC}{Probabilistic Error Cancellation}
\newacronym{spl}{SPL}{Sparse Pauli-Lindblad}
\newacronym{aces}{ACES}{Averaged Circuit Eigenvalue Sampling}

\begin{document}

\preprint{}

\title{Multi-Layer Cycle Benchmarking for high-accuracy error characterization}

\author{Alessio Calzona}
\affiliation{%
 IQM Quantum Computers, Georg-Brauchle-Ring 23-25, Munich, 80992,
Germany
}%
\author{Miha Papi\v{c}}
\affiliation{%
 IQM Quantum Computers, Georg-Brauchle-Ring 23-25, Munich, 80992,
Germany
}%
\affiliation{%
 Department of Physics and Arnold Sommerfeld Center for Theoretical Physics,\\
Ludwig-Maximilians-Universit\"at M\"unchen, Theresienstr. 37, 80333 Munich, Germany
}%
\author{Pedro Figueroa-Romero}\affiliation{%
 IQM Quantum Computers, Georg-Brauchle-Ring 23-25, Munich, 80992,
Germany
}%
\author{Adrian Auer}
\affiliation{%
 IQM Quantum Computers, Georg-Brauchle-Ring 23-25, Munich, 80992,
Germany
}%

\date{\today}

\begin{abstract}
Accurate noise characterization is essential for reliable quantum computation. Effective Pauli noise models have emerged as powerful tools, offering a detailed description of the error processes with a manageable number of parameters, which guarantees the scalability of the characterization procedure. However, a fundamental limitation in the learnability of Pauli eigenvalues impedes a full high-accuracy characterization of both general and effective Pauli noise models, thereby restricting {e.g.,} the performance of noise-aware error mitigation techniques. We introduce Multi-Layer Cycle Benchmarking (MLCB), an enhanced characterization protocol that improves the learnability associated with effective Pauli noise models by jointly analyzing multiple layers of Clifford gates. We show a simple experimental implementation and demonstrate that, in realistic scenarios, MLCB can reduce unlearnable noise degrees of freedom by up to $75\%$, improving the accuracy of sparse Pauli-Lindblad noise models and boosting the performance of error mitigation techniques like probabilistic error cancellation. Our results highlight MLCB as a scalable, practical tool for precise noise characterization and improved quantum computation. 
\end{abstract}

\maketitle

\section{Introduction}

A wide variety of characterization techniques, serving diverse purposes and spanning different ranges of applicability, have been developed to date. For instance, \gls{gst} provides a comprehensive method to determine individual errors associated with executing sets of quantum operations~\cite{Blume-Kohout2017Feb}, but its exponential scaling in sampling complexity limits its application to large systems~\cite{Nielsen2021}. On the other hand, holistic techniques like \gls{xeb}~\cite{Boixo2018, Chen2022} or scalable variants of \gls{rb}~\cite{Proctor2022, Hines2023, Hines2024} aim to benchmark large \glspl{qpu} by assessing their average quality.

Recently, considerable attention has been devoted to characterization schemes that lie between these extremes, offering greater detail than holistic techniques while still remaining scalable~\cite{vandenBerg2022Mar, Flammia2022, Pelaez2024, Mangini2024, Hockings2024}. This trend is also driven by the need to implement noise-aware error mitigation techniques~\cite{Temme2017Nov, vandenBerg2022Mar, Filippov2023, Kim2023} \add{and noise-aware decoding in the context of quantum error correction~\cite{Hockings2025Feb, Tiurev2023Sep}, both requiring} detailed knowledge of the noise associated with the parallel execution of instructions on large \glspl{qpu}. Achieving this without incurring the {prohibitively high} sampling complexity of \gls{gst} typically requires {implementing} averaging techniques {to} recast the original noise into {more manageable} forms. Equally important is the development of simpler, effective noise models that capture the most relevant features of the noise on the \add{entire} \gls{qpu} while featuring a manageable number of parameters~\cite{vandenBerg2022Mar, CarignanDugas2023, Mangini2024}.

A prominent example of {noise recasting} techniques is \gls{pt} applied to Clifford gates, which twirls noise into \rem{so-called} Pauli {noise}~\cite{Kern2005Jan,Wallman2016Nov, Hashim2021}. The latter is a type of stochastic noise {modeled by Pauli channels,} where only errors associated with any Pauli operator occur with a given probability. Incidentally, this makes \gls{pt} also a valuable error suppression tool, as it gets rid of a large portion of detrimental, \rem{so-called} {coherent} errors that get twirled into less detrimental, {incoherent} ones. This highlights the appeal and practical significance of characterizing Pauli noise~\cite{vandenBerg2022Mar, CarignanDugas2023}. Its relevance is further amplified by the possibility to develop and use effective and scalable versions of Pauli noise, where reasonable assumptions about noise locality can significantly reduce the number of parameters to be determined~\cite{vandenBerg2022Mar, Flammia2022, CarignanDugas2023}. 

Despite these advantages, the characterization of gate-dependent Pauli noise remains fundamentally limited by Pauli learnability constraints, as demonstrated in Ref.~\cite{Chen2023Jan} and confirmed experimentally~\cite{vandenBerg2022Mar, Ferracin2022, Hashim2021}. As in \gls{gst}~\cite{Nielsen2021, Merkel2013}, gauge freedom introduces ambiguities, making it impossible to determine certain parameters of Pauli channels independently from \gls{spam} errors. Specifically, Ref.~\cite{Chen2023Jan} shows that for a set of Clifford gates, the number of these unlearnable noise \gls{dof} in Pauli channels scales exponentially with the number of qubits $n$. In this context, \add{procedures based on \gls{cb} \cite{Erhard2019Nov, Flammia2020}} have emerged as convenient methods for accurately determining all learnable \gls{dof} of Pauli channels~\cite{Chen2023Jan}.

In this work, we specifically consider the characterization of scalable and effective Pauli noise models and present an enhanced version of \gls{cb} called \gls{mlcb}, which significantly increases the number of learnable \gls{dof} by leveraging the constraints inherent in the effective noise model. Unlike standard \gls{cb} implementations~\cite{vandenBerg2022Mar, CarignanDugas2023}, \gls{mlcb} simultaneously analyzes multiple sets of parallel Clifford gates, rather than characterizing each set in isolation. After providing a general framework for implementing \gls{mlcb}, we focus on the characterization of the IQM Garnet\textsuperscript{\tiny{TM}}~\cite{Abdurakhimov2024} quantum processor, which features $20$ superconducting qubits in a square topology. Our results show that \gls{mlcb} can achieve a remarkable $75\%$ reduction in the number of unlearnable \gls{dof} in practically relevant scenarios. Moreover, we experimentally highlight how \gls{mlcb}-enabled learnable \gls{dof} differ significantly from estimates derived through conventional low-accuracy approaches. Additionally, we numerically show how this enhanced learnability offered by \gls{mlcb} translates into more accurate characterization of effective noise models, focusing on the \gls{spl} noise model introduced in Ref.~\cite{vandenBerg2022Mar} and widely used to inform noise-aware error mitigation strategies~\cite{vandenBerg2022Mar, Filippov2023, Kim2023, Ferracin2022}. In this respect, we also investigate numerically how \gls{mlcb} reduces residual biases in \gls{pec}, a prominent error mitigation technique that requires precise noise characterization to achieve optimal results~\cite{Temme2017Nov, vandenBerg2022Mar, Govia2024, Ferracin2022}.

The rest of the paper is organized as follows: \add{Sections \ref{sec:2} and \ref{sec:spl} provide a short overview of the established \gls{cb} technique for Clifford gate layer characterization, discuss Pauli noise learnability constraints and briefly review scalable Pauli noise models, with a focus on the \gls{spl} model}. The novel \gls{mlcb} framework is presented in Section \ref{sec:mlcb}, followed by its experimental testing on the IQM Garnet\textsuperscript{\tiny{TM}} platform in Section \ref{sec:exp}. Finally, Sections \ref{sec:impro} and \ref{sec:em} numerically illustrate how \gls{mlcb} enhances characterization accuracy and improves the performance of noise-aware error mitigation techniques.

\section{Characterization of a Clifford layer}
\label{sec:2}

\add{A quantum circuit can be decomposed into a sequence of layers, where we define a layer as a set of instructions with non-overlapping qubit supports executed in parallel. Characterizing the performance of entire layers, rather than just individual gates, is crucial for assessing errors in practical applications, especially due to effects like spatial crosstalk where operations on some qubits also affect their neighbors. 

This work focuses on characterizing the noise of a $n$-qubit Clifford layer $C$ consisting of two-qubit entangling gates, whose execution typically represents the main source of errors. We model the noisy implementation of the layer $C$ as a map $\mu(U_C)[\rho] := (\phi(U_C) \circ \tilde \Lambda_C) [\rho]$, where $\phi(U_C)[\rho] := U_C \rho U_C^\dagger$ is associated with the unitary operator $U_C \in \mathbb{U}(2^n)$ and $\tilde \Lambda_C $ is a completely positive and trace-preserving noisy map. The latter can be simplified into an effective Pauli channel 
\begin{equation}
\label{eq:Lambda_C}
    \Lambda_C[\rho] = \sum_{\alpha} p_\alpha P_\alpha \rho P_\alpha^\dagger \quad(P_\alpha \in \mathbb{P}_n = \{I,X,Y,Z\}^{\otimes n}),
\end{equation}
by using \gls{pt}, which requires sandwiching the layer of interest $C$ with two layers of single-qubit Pauli gates \cite{Kern2005Jan, Wallman2016Nov, CarignanDugas2023, Hashim2021}. As adjacent layers of single-qubit Pauli operators can then be compiled together \cite{Wallman2016Nov}, the Pauli map $\Lambda_C[\rho]$ can be thus understood as the noise affecting the dressed Clifford layer $C$, i.e. the layer $C$ combined with a single layer of (random) Pauli gates. For more details about the twirling procedure, we refer the reader to App.~\ref{app:twirling} as well as to Refs. \cite{Kern2005Jan, Wallman2016Nov, CarignanDugas2023}. 

The task of characterizing the map $\Lambda_C[\rho]$ reduces to determining its Pauli eigenvalues $f_\beta$, associated with each Pauli operator $P_\beta$ as $\Lambda_C[P_\beta] = f_\beta P_\beta$ $\forall P_\beta \in \mathbb{P}_n$. Indeed, these Pauli eigenvalues, sometimes also referred to as Pauli fidelities \cite{Berg2022, Chen2023Jan, CarignanDugas2023}, can be directly related to the error probabilities in Eq.~\eqref{eq:Lambda_C} via the Walsh-Hadamard transform $p_\alpha = 2^{-n} \sum_\beta f_\beta (-1)^{\langle \alpha,\beta \rangle_{sp}}$, with $\langle \alpha, \beta \rangle_{sp}$ being the symplectic inner product of Pauli operators $P_\alpha$ and $P_\beta$, which is zero if they commute and one otherwise.}

\begin{figure}
    \centering
    \includegraphics[width=\linewidth]{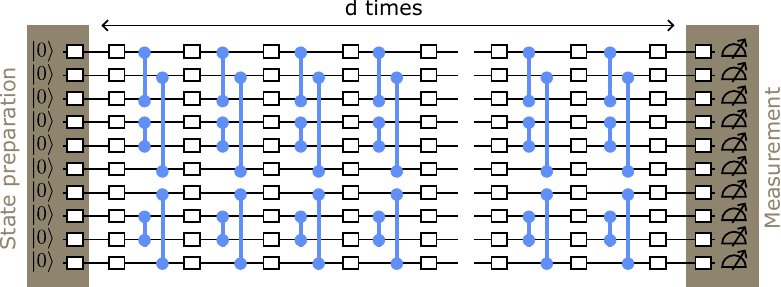}
    \caption{\textbf{Standard CB protocol}: In the state preparation stage (left brown area) an eigenstate of a Pauli operator is prepared. The Clifford layer $C$ subject to characterization (in this case, consisting of several parallel \cz gates shown in blue) is then implemented $d$ times, alternating it with layers of \add{single-qubit} gates (white boxes) which are needed to implement RC. This can be interpreted as the implementation of $d$ {dressed} Clifford layers. After a final layer of \add{single-qubit} gates which ensures that RC does not modify the unitary associated with the circuit, the measurement stage (right brown box), which also incorporates readout-twirling, provide us with the expectation value of a Pauli operator (typically the same {operator as} considered in the state preparation stage). 
    }
    \label{fig:SLCB}
\end{figure}

\add{Products of Pauli eigenvalues can be efficiently measured with high-accuracy \cite{Flammia2020} using \gls{cb}-structured circuits \cite{Erhard2019Nov}, as sketched in Fig.~\ref{fig:SLCB}. At first, a $+1$ eigenstate of the Pauli operator $P_\alpha$ is prepared; then the Clifford layer $C$ to be characterized is applied $d = km$ times (with $k, m \in \mathbb{N}_{>0}$); finally, the expectation value of $P_\alpha$ is measured. By implementing \gls{pt} targeted to both gate and readout errors \cite{Smith2021, Berg2022, vandenBerg2022Mar}, thus properly ensuring stochasticity of the latter \cite{Hashim2023, Beale2023Apr}, one can directly access the quantity 
\begin{equation}
A_\alpha \left(\prod_{\beta | P_\beta \in \mathcal{R}_\alpha^C}  f^C_\beta\right)^k.
\end{equation}
Here, the prefactor $A_\alpha$ fully accounts for SPAM errors (as constant offsets are eliminated by readout twirling) and $\mathcal{R}_\alpha^C$ is the orbit of the operator $P_\alpha$ under the application of $C$ \cite{CarignanDugas2023}. More specifically, we define it as the tuple 
\begin{equation}
\label{eq:orbit}
\mathcal{R}_\alpha^C = (P_\alpha, \phi(U_C)[P_\alpha], \; \dots\;, \phi(U_C)^{\circ (m-1)}[P_\alpha]),  
\end{equation}
where the orbit's cardinality $m\geq1$ is the smallest integer such that $\phi(U_C)^{\circ (m)}[P_\alpha] = P_\alpha$. Note that $m$ is upper bounded by the smallest positive integer $u$ such that $(U_C)^u = I$. By repeating the experiment for different values of $k$, it is thus possible to determine with multiplicative precision and in a SPAM-robust way \cite{Harper2019May, Flammia2020} the products of the Pauli eigenvalues associated with the operators belonging to each orbit $\mathcal{R}_\alpha^C$. Importantly, if an orbit features at least a pair of Pauli operators with the same support, its cardinality can be reduced by interleaving the layer $C$ with additional layers of specific single-qubit Clifford gates, which can be compiled together with the twirling layers \cite{Chen2023Jan, vandenBerg2022Mar}. In the remainder of this paper, we refer to this whole procedure as the standard CB protocol for measuring Pauli eigenvalues. 

The existence of orbits that cannot be reduced to a single element shows that some Pauli eigenvalues cannot be individually learned with high accuracy. This is not a flaw of the standard CB protocol but rather a manifestation of the general limited learnability of Pauli noise, stemming from the existence of a fundamental gauge freedom akin to the one encountered in the context of \gls{gst} \cite{Chen2023Jan, Nielsen2021}. Specifically, Ref.~\cite{Chen2023Jan} shows that out of the $4^n - 1$ \gls{dof} of general Pauli noise, the fundamentally unlearnable ones amount to $2^{n-c}$, where $c$ is the number of connected components of the pattern transfer graph based on the orbits analysis. As such, the limited learnability of Pauli noise also impacts alternative protocols developed to measure Pauli eigenvalues. Examples include tensor-network-based noise learning \cite{Mangini2024} and \gls{aces} \cite{Flammia2022, Hockings2024, Pelaez2024}. For instance, the \gls{aces} protocol can reconstruct all relevant Pauli eigenvalues associated with a local Pauli noise model only by arbitrarily fixing the gauge, e.g.~by assuming perfect state preparation \cite{Flammia2022, Hockings2024}.

A Pauli eigenvalue $f_\alpha$ being unlearnable does not imply that it cannot be estimated. For instance, this can be achieved using a modified CB-structured circuit with $d=1$, where an eigenstate of $P_\alpha$ is prepared and the expectation value of the conjugated Pauli operator $P_\beta = \phi(U_C)[P_\alpha]$ is measured \cite{McDonough2022, CarignanDugas2023, vandenBerg2022Mar}. Those alternative approaches, however, lack the SPAM-robustness and the multiplicative precision that the standard CB protocol provides for learnable quantities. In what follows, we will collectively refer to strategies used to estimate unlearnable Pauli eigenvalues as low-accuracy protocols, contrasted with the high accuracy of protocols like standard CB.}

\begin{table*}
\caption{\label{tab:conjugation}Conjugation of non-trivial Pauli operators in $\mathbb{P}_2$ under a single CZ (described by unitary $U_C$) and for a single CZ with extra S gates (unitary $\bar U_C$). Conjugated Pauli strings in bold are different from the original $P_\alpha$.}
\begin{ruledtabular}
\begin{tabular}{r|ccccccccccccccc}
$P_\alpha$& IX & IY & IZ & XI & XX & XY & XZ & YI & YX & YY & YZ & ZI & ZX & ZY & ZZ \\
$U_C P_\alpha U_C^{\dagger}$ & \textbf{ZX} & \textbf{ZY} & IZ &\textbf{XZ} &\textbf{ YY} & \textbf{YX }& \textbf{XI} & \textbf{YZ} & \textbf{XY} & \textbf{XX} & \textbf{YI} & ZI & \textbf{IX} & \textbf{IY} & ZZ  \\
$\bar U_C P_\alpha \bar U_C^{\dagger}$ & \textbf{ZY} & \textbf{ZX} & IZ &\textbf{YZ} & XX & XY & \textbf{YI} & \textbf{XZ} & YX & YY & \textbf{XI} & ZI & \textbf{IY} & \textbf{IX} & ZZ  \\
\end{tabular}
\end{ruledtabular}
\end{table*}

For the sake of concreteness, we conclude this section by discussing a simple example: the case of a single \cz gate performed on a $2$-qubit device. In this case, $U_C^2 =I$ limits the cardinality of the orbits to $2$, as it emerges also from Tab.~\ref{tab:conjugation}, which shows the conjugation of all non-trivial Pauli operators in $\mathbb{P}_2$ under a single \cz. Allowing for the usage of additional interleaved $S$ gates, which map $X\leftrightarrow Y$, there are $7$ Pauli eigenvalues which are directly learnable, i.e. associated with orbits with cardinality $1$. The remaining $8$ eigenvalues are not directly measurable but the standard \gls{cb} protocols allow to determine, with high accuracy, $8$ products of eigenvalue pairs. The latter are not all independent from each other since, for example, $f_{IY}f_{ZX} = (f_{IY}f_{ZY})(f_{IX}f_{ZX})/(f_{IX}f_{ZY})$. Still, they allow to impose a total of $6$ high-accuracy constraints on the parameters of the noise model, leaving us with only $2$ unlearnable \gls{dof}, which can be identified, for example, with Pauli eigenvalues $f_{IX}$ and $f_{XI}$. Those can only be estimated using low-accuracy protocols. More details concerning this simple example are provided in Appendix~\ref{app:ex_CZ}.

\section{Scalable Pauli characterization}
\label{sec:spl}

\add{Estimating all the $4^n-1$ parameters of the Pauli noise map associated with the execution of the layer $C$ on a $n$-qubit system is not possible at scale. It is therefore necessary to rely on simpler, effective Pauli models featuring a number of parameters that scales more favourably with $n$. A natural and established way to achieve this is to leverage the locality of the physical noise processes and the resulting sparsity of the model parameters. A prominent example here is the marginalization procedure, a key component of the Cycle Error Reconstruction (CER) scheme \cite{CarignanDugas2023} which has been successfully used to inform experimental realizations of noise-aware error mitigation technique \cite{Ferracin2022}. Another well-known scalable framework for characterizing Pauli noise is the so-called Sparse Pauli-Lindblad (SPL) model, introduced in Ref.~\cite{Berg2022} and successfully used to perform noise-aware error mitigation on large QPUs \cite{Kim2023, Fischer2024Nov}.}

\subsection{The SPL model}
\label{sec:spl}
\add{The limited learnability of Pauli noise directly also impacts effective and scalable models \cite{vandenBerg2022Mar, Fischer2024Nov, CarignanDugas2023}. Consequently, the implementation of \gls{mlcb} is beneficial, irrespective of the specific Pauli model under consideration. For the sake of concreteness, however, the remainder of this paper focuses primarily on the SPL model, which we introduce briefly below.

The \gls{spl} model assumes that} $\Lambda_C$ is generated by a Lindbladian whose jump operators are local and low-weight Pauli operators \cite{vandenBerg2022Mar}, that is $\Lambda_C =  e^{\mathcal{L}}$ with 
\begin{equation}
\mathcal{L}(\rho) = \sum_{\alpha\in\mathcal{K}} \lambda_\alpha^C (P_\alpha \rho P_\alpha - \rho).
\end{equation}
Here the set $\mathcal{K}$ consists of \add{all} Pauli strings with a support involving up to $w$ neighbouring qubits (with respect to the QPU's topology). \add{This also allows the model to capture crosstalk terms}.
\rem{To achieve a full characterization of the error model, all the model coefficients $\lambda^C_\alpha$ need to be determined.} Considering a maximum weight $w=2$ {---} a common assumption in the literature \cite{vandenBerg2022Mar, Kim2023, Filippov2024} and which we adopt throughout this paper {---} the number of model parameters associated with $\Lambda_C$ is given by $|\mathcal{K}| = 3n+9p$, \add{regardless of the number of gates in the layer}, with $n$ the number of qubits and $p$ the number of nearest-neighbour pairs. For a square topology, at scale, we have $p\sim 2n$ which leads to the linear scaling $|\mathcal{K}| \sim 21n$. 

The model parameters $\lambda_\alpha$ cannot be accessed directly but they are related to the Pauli eigenvalues \rem{, associated with the dressed Clifford layer $C$,} via
\begin{equation}
\label{eq:f<->lambda}
\log(f_\alpha)  = -2\sum_{\beta\in\mathcal{K}}  M_{\alpha \beta} \lambda_\beta,
\end{equation}
where $M_{\alpha \beta} = \langle \alpha,\beta \rangle_{sp}$ and we omitted the superscript $C$ for brevity. Because of limited learnability, sometimes only products of Pauli eigenvalues are known with high accuracy. \add{It is thus convenient} to define the full $|\mathcal{K}|$-dimensional vector of model parameters $\vec{\lambda}$ \footnote{Implicitly this means that we define an ordering of the set $\mathcal{K}$, which maps a Pauli string $\alpha$ to an integer $i \in \mathbb{N}$ so that the $i$-th element of $\vec{\lambda}$ (denoted by $\lambda_i$) is given by $\lambda^C_{\alpha}$.}. 
Similarly, we consider an $F$-dimensional vector $\vec \xi$ whose $j$-th element consists of a given product of eigenvalues, associated with the Pauli strings in a given set $\mathcal{X}_j$ (which typically corresponds to some orbit), i.e. $\xi_j = \prod_{\alpha\in \mathcal{X}_j} f_\alpha$. 
Then we can generalize Eq.~\eqref{eq:f<->lambda} to 
\begin{equation}
\label{eq:vec_f<>vec_lambda}
    \log(\vec \xi) = -2 \bar M \vec \lambda
\end{equation}
where the logarithm is applied element-wise and $\bar M$ is a rectangular $ F \times |\mathcal{K}|$ matrix whose $j$-th row reads $\bar M_{jk} = \sum_{\alpha \in \mathcal{X}_j} M_{\alpha k}$. If enough Pauli eigenvalues (or products thereof) are known such that the rank of $M$ is equal to $|\mathcal{K}|$, then Eq.~\eqref{eq:vec_f<>vec_lambda} can be inverted and the model parameters determined, typically via a non-negative least square fit
\begin{equation}
\label{eq:nnls}
    \vec\lambda(\vec \xi) = {\rm argmin}_{\lambda_k \geq 0}\;|| \bar M\vec\lambda + \log(\vec \xi)/2
||_2^2.
\end{equation}

Once all the model parameters $\vec{\lambda}$ are determined, the SPL model proves to be very useful. For example, it can be readily used to effectively amplify or reduce the circuit noise, enabling the execution of error mitigation techniques such as \gls{pec} \cite{Temme2017Nov,vandenBerg2022Mar}, tensor-network error mitigation (TEM) \cite{Filippov2023} and zero noise extrapolation \cite{Kim2023, Mari2021} with probabilistic amplification. Indeed, one can show that the (quasi)probabilistic implementation of the map \footnote{This can be done by randomly sampling circuits containing Pauli errors $P_k$ according to the (quasi)probability distribution defined by coefficients $w_k^{(\beta)}$ \cite{vandenBerg2022Mar}}
\begin{equation}
\Lambda_C^{(\beta)} (\rho) = \prod_{k\in \mathcal{K}} \left( w_k^{(\beta)} \cdot + (1 -w_k^{(\beta)}) P_k \cdot P_k\right) \rho
\end{equation}
before the layer $C$ effectively scales the overall noise as $\Lambda_C \circ \Lambda_C^{(\beta)} = e^{\beta \mathcal{L}}$, provided that $2w_k^{(\beta)} = 1-\exp[-2(\beta-1)\lambda_k]$ \cite{McDonough2022}. 

\subsection{Effects of limited learnability on the SPL model}

Let us consider the characterization of a Clifford layer $C$ composed of parallel \cz gates {---} a scenario of significant practical relevance \cite{Abdurakhimov2024, Google2023Feb, McKay2023}. In order to fit the corresponding \gls{spl} model, it is sufficient to know~\cite{vandenBerg2022Mar} (i) all the $3n$ eigenvalues associated with \add{weight-one} Pauli operators and (ii) all products of pairs of $f_\alpha f_\beta$ where $\alpha\in\mathcal{K}$ is a weight-two Pauli string and $P_\beta$ is the conjugation of $P_\alpha$ under the layer $C$ \add{and, possibly, additional single-qubit Clifford gates} \footnote{Indeed, it is required that each one of the two elements of $\beta$ in the support of $\alpha$ is either the identity or equal to the corresponding element of $\alpha$. This can be always ensured by composing the layer $C$ with a second Clifford layer, consisting only of \add{single-qubit} gates\rem{in the same spirit as the interleaved \gls{cb}}~\cite{vandenBerg2022Mar}}. While the \rem{combination of} standard \rem{and interleaved} \gls{cb} protocol provides us with high-accuracy values for all the products of eigenvalues required in (ii), condition (i) is highly impacted by the limited learnability of Pauli noise and some eigenvalues will need to be estimated using alternative, low-accuracy approaches.

In the previous section, we demonstrated that a single, isolated CZ gate has two unlearnable \gls{dof}, which can be conveniently associated with the \add{weight-one} Pauli eigenvalues $f_{IX}$ and $f_{XI}$. Leveraging this fact, along with condition (i), we can immediately compute the number of unlearnable \gls{dof} for an \gls{spl} characterization of a layer containing $n_g$ parallel \cz gates, which totals $2 n_g$. These can also be conveniently associated with $2 n_g$ unlearnable \add{weight-one} eigenvalues, each one of them associated with a single $X$ operator on one qubit belonging to the layer's support. 

\add{Several methods have been proposed to estimate these unlearnable quantities. One approach is the non-SPAM-robust $d=1$ CB variant mentioned earlier. Another common—but theoretically unjustified—strategy relies on assuming symmetries among the Pauli eigenvalues. For instance, in the case of a single CZ gate, one might assume that unlearnable eigenvalues are equal to their conjugates under the action of  $C$, such as $f_{XI} = f_{XZ}$. This allows the estimation of individual unlearnable eigenvalues via $f_{XI} = \sqrt{f_{XI}f_{XZ}}$, where the product on the right-hand side is learnable. However, since this assumption lacks rigorous justification, the method can introduce significant errors and yields low accuracy.}

Regardless of the chosen strategy for addressing limited learnability, once conditions (i) and (ii) are satisfied, all information about the (products of) eigenvalues can be encapsulated in the vector $\vec\xi$. It is useful to represent $\vec\xi$ as a concatenation $\vec\xi = (\vec \Xi, \vec \zeta)$, where $\vec \Xi$ contains only learnable (products of) eigenvalues, and $\vec \zeta$ comprises unlearnable ones. At this stage, the full-rank matrix $\bar M$ can be constructed, enabling the \gls{spl} model parameters $\vec{\lambda}$ to be determined by inverting Eq.~\eqref{eq:vec_f<>vec_lambda}. As shown in Refs.~\cite{Chen2023Jan, vandenBerg2022Mar} and further discussed in Sec.\ref{sec:impro}, the limited accuracy associated with $\vec \zeta$ impacts the overall precision of the scalable error characterization.

\begin{figure}
    \centering
    \includegraphics[width=.7\linewidth]{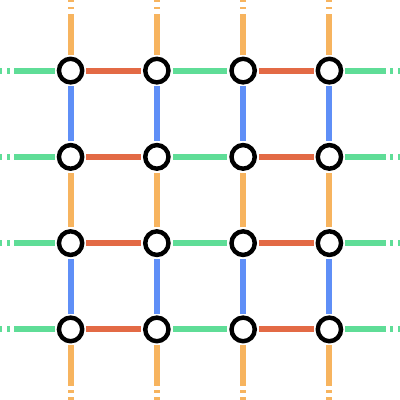}
    \caption{A \gls{qpu} with square topology and four layers of parallel \cz gates, identified with blue, green, red and orange colors, that covers all nearest-neighbour connections between the qubits.}
    \label{fig:square}
\end{figure}

So far, we have focused on the characterization of a single layer $C$. However, in realistic scenarios, multiple distinct layers must often be addressed. A particularly relevant example is depicted in Fig.~\ref{fig:square}, which illustrates a large $n$-qubit QPU with a square topology and four layers of parallel \cz gates, color-coded as blue (B), green (G), red (R), and orange (O). These four layers collectively cover all nearest-neighbor connections on the \gls{qpu}. Combined with arbitrary \add{single-qubit} gate layers, they enable the construction of non-trivial circuits that facilitate rapid entanglement distribution across the whole \gls{qpu} \cite{Arute2019}. Such dense configurations are useful for a variety of applications, including simulating the dynamics of lattice models \cite{Kim2023} and implementing variational quantum algorithms \cite{OBrien2022}.

In the limit of large $n$, the number of \cz gates per layer scales linearly as $n_g \sim 2 n$, as does the number of \gls{spl} parameters per layer, $|\mathcal{K}| \sim 21n$. Characterizing all four layers requires determining a total of $\sim 84 n$ parameters. If each layer is characterized independently, the total number of unlearnable quantities amounts to $2 n_g \sim  4n$. Unlike the (non-scalable) full characterization of Pauli noise—where the unlearnable \gls{dof} constitute an exponentially small fraction of the learnable ones \cite{Chen2023Jan}—this example shows that for large $n$ and considering the \gls{spl} model, the ratio of unlearnable \gls{dof} to total parameters approaches a constant value of around $4/84\sim4.8\%$. As we will discuss in Sections \ref{sec:impro} and \ref{sec:em}, this small fraction is enough to significantly hinder the overall quality of error characterization and the effectiveness of noise-aware error mitigation techniques.

\subsection{Number of required experiments}

Compared to strategies based on depth-one circuits, approaches relying on symmetry assumptions have the significant advantage of not requiring additional experiments. In these cases, the entire experimental runtime is indeed allocated to executing the standard \gls{cb} protocol. For the sake of concreteness, we still focus on \gls{qpu}s with a square lattice topology, as illustrated in Fig.~\ref{fig:square}. In this scenario, it has been shown that the (products of) eigenvalues associated with all orbits $\mathcal{R}^C_\alpha$ for $\alpha\in\mathcal{K}$ can be measured using the standard \gls{cb} protocol with just eleven \gls{cb} instances, a single instance referring to given state preparation and measurement stages and potential interleaving of extra single-qubit Clifford gates~\cite{vandenBerg2022Mar} \footnote{This result is not limited to square lattices but applies more generally to topologies where qubits can be ordered such that no qubit is directly connected to more than two of its predecessors.}. \rem{As for the interleaved \gls{cb} instances, building on the simple single \cz gate example, it can be shown that just two additional instances of interleaved \gls{cb}—where extra S gates are added before each \cz gate—are required to capture all the necessary additional (products of) eigenvalues. Consequently, a single layer of parallel \cz gates can be fully characterized using a total of $11$ \gls{cb} instances. Here, a single instance refers to a CB experiment with a fixed initial Pauli state and measurement basis}. For the four layers shown in Fig.~\ref{fig:square}, this approach requires a total of $44$ \gls{cb} instances.

This number can be further reduced to $16$ instances (four per layer) by generalizing the \gls{pt} procedure to include S gates in the twirling set. This extension symmetrizes the $X$ and $Y$ components of the noise, reducing the amount of information that needs to be extracted and thereby decreasing the experimental runtime \cite{Berg2023}. It is important to emphasize that the discussion here focuses solely on the number of unique \gls{cb} instances required—an essential aspect of implementing the procedure efficiently \cite{Berg2023}. However, this does not address the sampling complexity of the characterization strategies, which is also an important factor in determining the overall runtime. A more comprehensive discussion of the sampling complexity associated with Pauli noise characterization can be found in Ref.~\cite{Chen2023a}.

\section{Multi-layer CB}
\label{sec:mlcb}

\begin{figure}
    \centering
    \includegraphics[width=\linewidth]{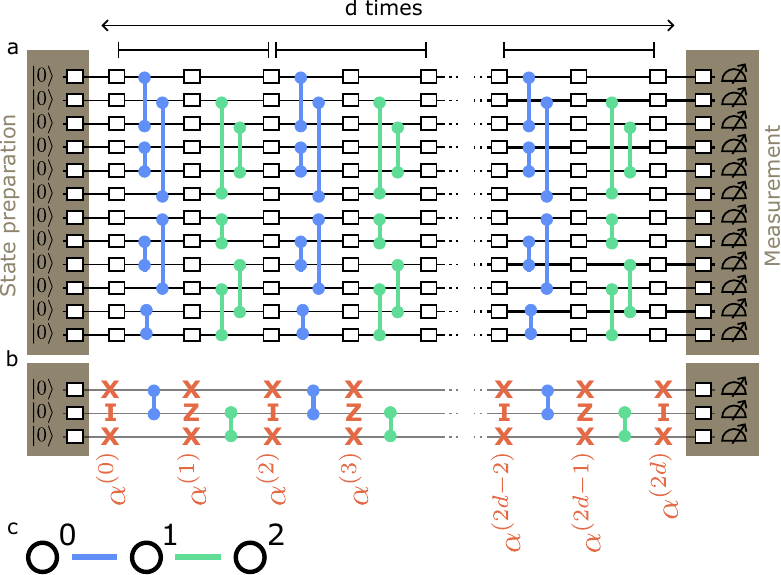}
    \caption{\textbf{\gls{mlcb} protocol}: (a) The structure is analogous to the one shown in Fig.~\ref{fig:SLCB} with the important difference that, here, the ``building block'' which is repeated $d$ times is not a single Clifford layer but a combination of more layers (in this specific case two, shown in blue and green). Single-qubit gates used for twirling are indicated in white. (b) Simple scenario where \gls{mlcb} allows one to measure with high accuracy the product $f^B_{XIX}f^G_{XZX}$ of two eigenvalues, belonging to two different layers, which cannot be accessed with conventional CB protocols. For the sake of simplicity, single-qubit gates used for twirling are now explicitly shown. (c) Simple system consisting of three qubits and two layers.}
    \label{fig:MLCB}
\end{figure}

In realistic scenarios, quantum circuits typically involve more than a single (dressed) Clifford layer, requiring the characterization of an entire set $\mathcal{C}$ of distinct layers. The {conventional} approach to this problem involves characterizing each unique layer individually, following the procedure outlined earlier \cite{Kim2023, vandenBerg2022Mar}.
In this section, we propose augmenting the {conventional} approach with the novel \gls{mlcb} strategy that leverages circuits retaining the fundamental \gls{cb} structure but relying on building blocks that consist of combinations of different layers, picked from the set $\mathcal{C}$. A simple yet illustrative example is shown in Fig.~\ref{fig:MLCB}(a), where a combination of two layers, depicted in blue and green, is repeated $d$ times. Note that each layer is independently twirled, as shown by the white boxes indicating single-qubit twirling gates. 

\add{To properly describe the \gls{mlcb} protocol, it is convenient to introduce the definition of a multi-layer orbit associated with the pair of layers $L_1, L_2 \in \mathcal{C}$. It generalizes Eq.~\eqref{eq:orbit} and reads 
    \begin{equation}
        \label{eq:ml_orbit}
        \mathcal{R}_\alpha^{L_1 L_2} = \left(P_{\alpha^{(0)}}, P_{\alpha^{(1)}},  \dots,  P_{\alpha^{(m-1)}}\right)
    \end{equation}
    where 
    \begin{equation}
    \label{eq:pauli_strings}
        P_{\alpha^{(j)}} = \phi(\bar U_{L_1L_2}^{(j)})[P_\alpha]
    \end{equation}
    and the unitary 
    \begin{equation}
        \label{eq:Um}
        \bar U_{L_1L_2}^{(l)} = \begin{cases}
    (U_{L_1}U_{L_2})^{l/2}& \text{ if } l\text{ is even }, \\
    U_{L_1}(U_{L_2}U_{L_1})^{(l-1)/2} &  \text{ if } l\text{ is odd }, 
        \end{cases}
    \end{equation}
    captures the effect of two alternating layers $L_1, L_2\in \mathcal{C}$, associated with unitaries $U_{L_1}$ and $U_{L_2}$. The cardinality $m\geq2$ of the multi-layer orbit $ \mathcal{R}_\alpha^{BG}$ is the smallest even integer such that $\phi(\bar U_{L_1L_2}^{(m)})[P_\alpha] = P_\alpha$.} \add{Analogously to the standard \gls{cb} protocol, by preparing a $+1$ eigenstate of the operator $P_\alpha$ and measuring its expectation value after the implementation of $m/2$ pairs of twirled layers, the \gls{mlcb} protocol can thus measure in a SPAM-robust way and with multiplicative precision the product of the Pauli eigenvalues 
    \begin{equation}
        \mathcal{F}_\alpha^{L_1L_2} = \prod_{j=0}^{m/2-1} f^B_{\alpha^{(2j)}} f^G_{\alpha^{(2j+1)}},
    \end{equation}
        where the Pauli strings are defined according to Eq.~\eqref{eq:pauli_strings}. }

    In the remainder of this Section, we demonstrate how this technique can be used to boost the learnability of Pauli noise models by focusing at first on the characterization of a simple system and then discussing its generalization to more complicated and relevant scenarios. 

\subsection{Simple example}
\label{sec:simple_ex}
	The simple setup shown in Fig.~\ref{fig:MLCB}(c) consists of three qubits arranged in an open chain and numbered from $0$ to $2$. The goal is to characterize the two depicted layers, which correspond to a single \cz between qubits $0$ and $1$ (blue layer) and between qubits $1$ and $2$ (green layer). The {conventional} approach, described in the previous Section, prescribes to focus on each layer individually and perform \gls{cb} protocols to determine with high accuracy all the (products of) eigenvalues related to the orbits that involve the Pauli strings $\alpha\in \mathcal{K}$. Since each layer consists of a single CZ gate, there are four unlearnable \gls{dof}, which can be identified with the set of weight-one eigenvalues
	\begin{equation}
	    \label{eq:unlearnable}
	    \left\{f^B_{XII}, f^B_{IXI}, f^G_{IXI}, f^G_{IIX}\right\}.
	\end{equation}
    Here we use the notation $f^L_\alpha$ for the Pauli eigenvalues, where $\alpha$ is a Pauli string (ordered according to the increasing qubit indexes) and $L\in\{B,G\}$ indicates the associated layer, i.e. the blue (B) one or the green (G) one. To complete the characterization task,
    all the eigenvalues in \eqref{eq:unlearnable} must be estimated using low-accuracy methods, such as {symmetry assumptions} or unit-depth variants of \gls{cb} protocols.

   \add{Notably, the \gls{mlcb} experiment shown in Fig.~\ref{fig:MLCB}(b) significantly advances the characterization process by enabling the high-accuracy determination of an additional \gls{dof}. This is accomplished by analyzing the orbit $\mathcal{R}_{XIX}^{BG} = (P_{XIX}, P_{XZX})$, which yields a precise estimate of the product of individually unlearnable Pauli eigenvalues:
    \begin{equation}
        \mathcal{F}^{BG}_{XIX} = f^B_{XIX} f^G_{XZX}.
    \end{equation}
    These eigenvalues correspond to high-weight, non-local Pauli operators, which would be unrelated to the unlearnable \gls{dof} in Eq.\eqref{eq:unlearnable} in the context of a general Pauli noise model. Within the SPL framework, however, not all Pauli eigenvalues are independent from each other and, as we demonstrate in Appendix \ref{app:equivalence}, the knowledge of $\mathcal{F}^{BG}_{XIX} $ provided by \gls{mlcb} is equivalent to the knowing the ratio $\mu_1^{BG} \equiv {f^B_{XII}}/{f^G_{IIX}}$ between two unlearnable weight-one eigenvalues in Eq.~\eqref{eq:unlearnable}. Specifically, this relationship is expressed as
    \begin{equation}
	\label{eq:equiv_2}
	   \mu_1^{BG} \equiv \frac{f^B_{XII}}{f^G_{IIX}} = \mathcal{F}^{BG}_{XIX} \, \frac{(f^G_{IZI})}{(f^B_{IIX})(f^G_{XZI})(f^G_{IZX}f^G_{IIX})}
	\end{equation}
     where all the terms in round brackets are learnable and can be determined with high-accuracy using conventional \gls{cb} protocols. The non-trivial structure of Eq.~\eqref{eq:equiv_2} arises because the SPL model accounts for crosstalk, which introduces correlated errors also between neighboring qubits $0$ and $1$.

     Building on this insight, we generalize the approach, presenting a systematic method to identify relevant multi-layer orbits. This allows \gls{mlcb} to impose high-accuracy constraints on ratios of otherwise unlearnable weight-one Pauli eigenvalues for arbitrary combinations of Clifford layers.}

	\subsection{Generalization to two arbitrary layers}

\begin{figure*}
    \centering
    \includegraphics[width=\linewidth]{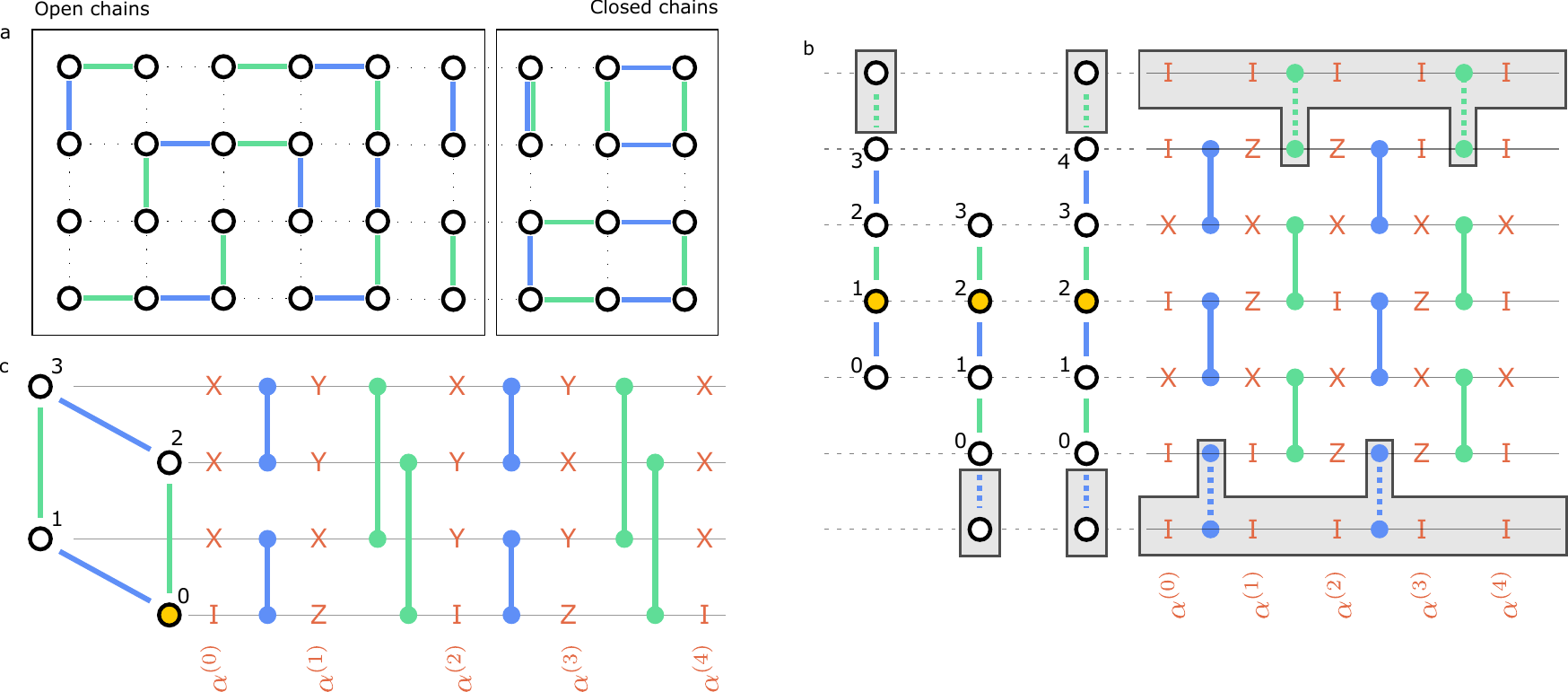}
    \caption{(a) An example of several disconnected graphs $G_i$ obtained by considering edges associated with two layers, blue and green, of parallel \add{two-qubit} gates. The graphs can be divided into open and closed chains, as shown by the black rectangles. (b) Analysis of open chains showing the conjugations of specific Pauli strings $\alpha^{(m)}$ (with $m=0, \dots 4$) under four subsequent layers, in the pattern BGBG, that constitute the building block of \gls{mlcb} protocols. Different chains, featuring four and five qubits, are depicted on the left (each with a specific qubit numbering). By looking at the corresponding qubits (and gates) on the right side one can identify the circuit blocks that allow to measure specific products of eigenvalues. The latter can be then used to determine the ratios $\mu_q^{BG}$ of unlearnable eigenvalues associated with the qubit $q$ (highlighted in yellow). Gray-shaded areas shows scenarios in which that the presence of additional qubit is trivial. (c) Analysis of closed chains consisting of four qubits. As for the previous panel, the circuits shows how MLCB can measure a product of four eigenvalues, which can then be used to determine the ratio $\mu_q^{BG}$ (with $q=0$ highlighted in yellow).}
    \label{fig:chains}
\end{figure*}

We consider the characterization of two generic $n$-qubit (dressed) layers, identified with blue (B) and green (G) colors. We begin by deriving conclusions applicable to general layers consisting of parallel \add{two-qubit} Clifford gates and then proceed by specifically focusing on the relevant case of layers consisting of parallel \cz gates on a \gls{qpu} with square lattice topology. 

A useful way to visualize the problem at hand is to represent each \add{two-qubit} gate as a colored edge connecting the two qubits in its support. Those edges collectively form several disconnected graphs $G_i$ over the \gls{qpu}. An example of this visualization, applied to a square topology, is shown in Fig.~\ref{fig:chains}(a). Since the Clifford layers consist of {parallel} \add{two-qubit} gates, a qubit cannot simultaneously belong to two edges of the same color. Consequentially, each graph $G_i$ is a one-dimensional chain, either open or closed, composed of edges with alternating colors. Moreover, closed chains necessarily consist of an even number of qubits. This simple structure, which arises independently of the \gls{qpu} topology, significantly simplifies the analysis of \gls{mlcb} protocols, as each graph $G_i$ can be analyzed independently and in parallel. In particular, for each graph $G_i$ we thus need to identify the Pauli string $\alpha$, and the associated multi-layer orbit $\mathcal{R}_\alpha^{BG}$, which allows us to learn \gls{dof} which would have been otherwise unlearnable.

In the remainder of this Section, we specifically focus on Clifford layers consisting of parallel \cz gates on a square topology. This limits the cardinality of the multi-layer orbits to $4$, as $(U_BU_G)^2 = I$. In this context, we are able to prove that, for every qubit $q$ in the bulk of a graph $G_i$, the \gls{mlcb} protocol enables the high-accuracy determination of one additional \gls{dof} that would have been unlearnable using conventional \gls{cb} methods alone. In particular, this \gls{dof} can be identified with the ratio $\mu_q^{(BG)}$ of two individual unlearnable eigenvalues
\begin{equation}
	\mu_q^{(BG)} = \frac{f^B_\beta}{f^G_\gamma}.
\end{equation}
Here the weight-one Pauli string $\beta$ ($\gamma$) consists only of identities $I$ with the exception of a single $X$ associated to the qubit connected to $q$ via the blue (green) edge. This definition explains the usage of the symbol $\mu_1^{(BG)}$ in Eq.~\eqref{eq:equiv_2}, where qubit number $1$ is the only one in the bulk of the graph shown in Fig.~\ref{fig:MLCB}(c).

Remarkably, all of those \gls{dof} can be accessed by preparing Pauli strings $\alpha$ consisting solely of $I$ and $X$ operators, regardless of the specific qubit $q$ or graph $G_i$. Consequently, by preparing a $+1$ eigenstate of $X^{\otimes n}$ and measuring in the $X$ basis, it becomes possible to determine all the eigenvalue ratios $\mu_q^{(BG)}$ simultaneously, with a single \gls{mlcb} execution. This parallelization introduces only a limited overhead compared to
the conventional procedure. Together with the details in Appendix~\ref{app:equivalence}, the remainder of this section provides a concrete and operational framework for implementing \gls{mlcb} in relevant scenarios, highlighting the protocol's capability to unlock additional \gls{dof} in an efficient way.

\subsubsection{Open chains}
Let us start with the analysis of open chains, some of which are depicted in the left rectangle of Fig.~\ref{fig:chains}(a). Open chains consisting of just two qubits feature a single layer (and have no bulk), meaning that \gls{mlcb} cannot be implemented. As for open chains with three qubits, they have been already fully covered in Section \ref{sec:simple_ex}. We therefore proceed by focusing on open chains featuring a total of four qubits. To analyze them within the \gls{mlcb} framework, it is necessary to repeatedly implement the minimal four-layer block with a BGBG pattern, as shown in Fig.~\ref{fig:chains}(b). If the open chain features two blue \cz gates, Fig.~\ref{fig:chains}(b) shows how \gls{mlcb} can directly measure the product of {four} eigenvalues $o_4 = f^B_{XIXI}f^G_{XZXZ}f^B_{XIXZ}f^G_{XZXI}$. In Appendix \ref{app:equivalence}, we prove it to be equivalent to the ratio $\mu_{1}^{(BG)}$ between two unlearnable eigenvalues associated with the bulk qubit $1$. Similarly, if the open chain features two green \cz gates, \gls{mlcb} can directly measure $o'_4 = f^B_{IXIX}f^G_{IXZX}f^B_{ZXIX}f^G_{ZXZX}$, which we prove to be equivalent to the ratio $\mu_{2}^{(BG)}$. By exploiting inversion symmetry, analogous analyses can be performed to determine the ratios of unlearnable eigenvalues associated with the other qubits in the bulk of four-qubit chains. 

As for open chains consisting of five qubits, as shown in Fig.~\ref{fig:chains}(b), there are three qubits in the bulk. The ratios $\mu_q^{(BG)}$ associated with qubits $q=1,3$ can be readily determined by following the strategies discussed above for the four-qubit chains, as the relevant Pauli strings $\alpha^{(m)}$ always have an identity $I$ operator on the qubit at the opposite side of the chain (see the gray-shaded areas). By contrast, to determine the ratio associated with the central qubit, all the five qubits are involved. In particular, \gls{mlcb} can directly measure the product $o_5 = f^B_{IXIXI}f^G_{IXZXZ}f^B_{ZXIXZ}f^G_{ZXZXI}$, which we then prove in Appendix to be equivalent to $\mu_2^{(BG)}$. Interestingly, even if more qubits were present (see the gray-shaded areas), the relevant Pauli strings $\alpha^{(m)}$ always feature operators different from the identity $I$ on no more than five qubits. As a result, when analyzing longer open chains, we can readily determine the ratio $\mu_q^{(BG)}$ for each qubit in the bulk by resorting to the \gls{mlcb} strategies detailed so far.

\subsubsection{Closed chains}
The shortest closed chain consists of just two qubits, both residing in the bulk, each paired by a \cz gate in both layers (see in the upper left corner of the ``Closed chains'' rectangle in Fig.~\ref{fig:chains}(a)). In this straightforward case, the \gls{mlcb} approach directly provides us with two products of eigenvalues, $c_2 = f^B_{IX} f^G_{ZX}$ and $c'_2 = f^B_{XI} f^G_{XZ}$, which are equivalent to the two ratios $\mu_0^{(BG)}$ and $\mu_1^{(BG)}$. The analysis of chains consisting of four qubits is more involved. In the case shown in Fig.~\ref{fig:chains}(c), \gls{mlcb} allows one to measure $c_4 = f^B_{IXXX} f^G_{ZXYY} f^B_{IYYX} f^G_{ZYXY}$, a product of eigenvalues that is equivalent to the ratio $\mu_q^{(BG)}$ of unlearnable eigenvalues associated with qubit $q=0$. By rotating and/or mirroring the chain, one can systematically derive the ratios for the other three qubits $q$ in the bulk of the chain. Detailed derivation is provided in Appendix \ref{app:equivalence}. For longer chains involving six or more qubits, the results derived for open chains can be directly applied. This is because the latter never involves more than five qubits at a time, regardless of the chain length. A clear example of this can be seen from Fig.~\ref{fig:chains}(b) by identifying (as a single qubit number $5$) the two gray-shaded qubits, effectively showing how to use \gls{mlcb} to determine the ratio $\mu_q^{(CB)}$ associated with each qubit in a bulk of a six-qubit closed chain. 

\subsection{More than two layers} 
The approach detailed so far can be readily generalized to scenarios featuring more than two layers. Let us focus on a specific qubit $q$, which belongs to the support of $l$ layers in the set $\{L_1, \dots, L_l\}$. By simply implementing the procedure outlined before, for each pair of layers $(L_i L_j)$ we can thus determine the ratio $\mu_q^{(L_iL_j)}$, which would have been unlearnable if only standard \gls{cb} were used. Unfortunately, not all ratios $\mu_q^{(L_iL_j)}$ are independent from each other. Specifically, the knowledge of the $l-1$ ratios $\mu_q^{(L_iL_{i+1})}$ (for $i=1, \dots, l-1$) is enough to reconstruct all the remaining ones, meaning that \gls{mlcb} provides us with $l-1$ DOF that would have been unlearnable otherwise. 

Let us discuss the concrete effect of \gls{mlcb} on the highly structured example in Fig.~\ref{fig:square}(a), where every qubit $q$ in the bulk of the \gls{qpu} belongs to the support of each one of the four layers, blue (B), green (G), red (R) and orange (O). By performing \gls{mlcb} on three pairs of layers, we can determine three independent ratios per qubit, say $\mu_q^{(BG)}$, $\mu_q^{(GR)}$, and $\mu_q^{(RO)}$. In the limit of large $n$, this brings down the total number of unlearnable \gls{dof} that needs to be determined with low-accuracy methods from $\sim 4 n$ to $\sim n$. Importantly, this $75\%$ reduction of unlearnable DOF comes with a rather small experimental overhead, around $7\%$ more circuit runs, since it requires to perform only $3$ extra \gls{mlcb} experiments on top of the $44$ \gls{cb} instances required in any case by the standard approach. This shows how \gls{mlcb} can be beneficial, especially in the highly relevant context of characterizing errors during the execution of circuits with few dense layers of \cz gates. 

\section{Experimental implementation}
\label{sec:exp}

\begin{figure*}
    \centering
    \includegraphics[width=1\textwidth]{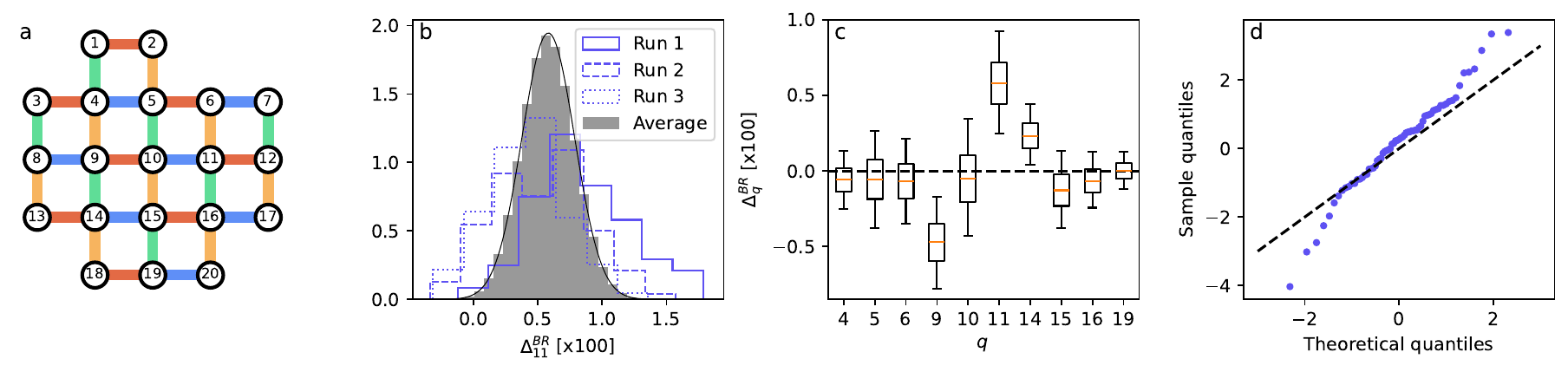}
    \caption{(a) Layout of the IQM Garnet\textsuperscript{\tiny{TM}} \cite{Abdurakhimov2024} quantum processor, highlighting the four characterized layers: blue (B), green (G), red (R), and orange (O). (b) Bootstrapped distributions of the difference $\Delta_{11}^{BR}$ between products of high-weight Pauli eigenvalues, as measured by \gls{mlcb}, for three consecutive experimental runs (purple) and their average (gray). (c) All measured differences $\Delta_q^{BR}$ for the BR layer pair. (d) Q-Q plot of the ratios $\Omega_q^{L_1L_2}$ (defined in Eq.~\eqref{eq:ratio_Omega}) for all layer pairs $\{\rm BR, BO, BG, RO, RG, OG\}$ and a reference standard normal distribution (black dashed line).}
    \label{fig:exp}
\end{figure*}

\add{
We experimentally validate the \gls{mlcb} protocol on IQM Garnet\textsuperscript{\tiny{TM}}, a 20-qubit superconducting quantum computer \cite{Abdurakhimov2024}. Our objectives are twofold: first, to use \gls{mlcb} to obtain high-accuracy data inaccessible to conventional \gls{cb} protocols; and second, to test the common—but theoretically unjustified—assumption of symmetry between unlearnable Pauli eigenvalues. The protocol is robust against both temporal drifts in system noise and deviations from the underlying effective Pauli noise model.

We focus on the four gate layers depicted in Fig.~\ref{fig:exp}(a), which pairwise define open chains of various lengths. Under the \gls{spl} model, for each qubit $q$ in the bulk of a chain formed by layers $L_1$ and $L_2$, \gls{mlcb} enables us to learn the ratio $\mu_q^{(L_1L_2)}$ of weight-one eigenvalues, which are unlearnable using conventional \gls{cb}. This is achieved by directly measuring products of non-local and high-weight Pauli eigenvalues $\mathcal{F}_{\alpha(q)}^{L_1L_2}$, without assuming a specific Pauli noise model. Importantly, the state-preparation and measurement stages can be chosen such that the same experimental \gls{mlcb} data also yield the product $\mathcal{F}_{\beta(q)}^{L_1L_2}$, where the initial Pauli operators are related by conjugation, $P_{\beta(q)} = \phi(U_{L_1})[P_{\alpha(q)}]$. Conventional \gls{cb} protocols cannot resolve these quantities individually; they can only measure the overall product $f = \mathcal{F}_{\alpha(q)}^{L_1L_2} \mathcal{F}_{\beta(q)}^{L_1L_2}$. Without \gls{mlcb}, estimating the individual terms requires low-accuracy methods, such as assuming $\mathcal{F}_{\alpha(q)}^{L_1L_2} \approx \mathcal{F}_{\beta(q)}^{L_1L_2} \approx \sqrt{f}$. We directly test this assumption by defining their difference as:
\begin{equation}
\label{eq:delta_q_l1l2}
\Delta_q^{L_1L_2} = \mathcal{F}_{\alpha(q)}^{L_1L_2} - \mathcal{F}_{\beta(q)}^{L_1L_2}.
\end{equation}
A measured value of $\Delta_q^{L_1L_2}$ statistically incompatible with zero provides direct, Pauli model-independent, and time-drift-robust evidence that the symmetry assumption is violated and that \gls{mlcb} yields information beyond the reach of conventional methods. This approach is inherently robust against temporal variations, as differences are computed from the same dataset, eliminating potential systematic errors from comparing distinct datasets obtained under different noise conditions.

For each pair of layers in Fig.~\ref{fig:exp}(a), we perform three back-to-back \gls{mlcb} experiments. While each measured difference $\Delta_{q}^{L_1L_2}$ is inherently drift-robust, this approach also allows us to probe the temporal stability of $\Delta_{q}^{L_1L_2}$ while accumulating additional statistics. Fig.~\ref{fig:exp}(b) exemplifies this for $\Delta_{11}^{BR}$, associated with qubit $q=11$ and the blue (B) and red (R) layers. After confirming mutual consistency among the differences from each run, we average the data to improve statistical precision. The results of each experiment are bootstrapped 100 times with respect to both shots and random Pauli twirling instances, yielding the three purple histograms in Fig.~\ref{fig:exp}(b) (one per run). The data from the three runs are then averaged, resulting in the gray histogram, which is well-described by a Gaussian (fitted black line). We compute the ratio:
\begin{equation}
\label{eq:ratio_Omega}
\Omega_{q}^{L_1L_2} = \text{Mean}(\Delta_{q}^{L_1L_2})/\text{Std}(\Delta_{q}^{L_1L_2}),
\end{equation}
which for $\Delta_{11}^{BR}$ is $\Omega_{11}^{BR} \sim 2.84$, corresponding to a probability of $\Delta_{11}^{BR}$ not being compatible with zero of approximately $99.5\%$.

As detailed in Appendix~\ref{app:exp}, we extend this analysis to all measured $\Delta_q^{L_1L_2}$. The box plot in Fig.~\ref{fig:exp}(c) summarizes the results for the BR layer pair, with whiskers indicating the 95\% confidence interval derived from bootstrapping, showing another difference, $\Delta_{9}^{BR}$, which is most-likely not compatible with zero. We verify that the bootstrapped distributions are Gaussian (with the exception of one manifestly non-Gaussian distribution, $\Delta_{5}^{OR}$, which we then exclude from the following analysis). The Q-Q plot in Fig.~\ref{fig:exp}(d) shows the distribution of all the ratios $\Omega_{q}^{L_1L_2}$. With 67 ratios, the theoretical quantiles on the $x$ axis (computed assuming a normal distribution and Filliben’s estimate) lie in the interval $(-2.32, 2.32)$, while the experimental $\Omega_{q}^{L_1L_2}$ on the $y$ axis span a much larger range $(-4.03, 3.39)$. This confirms the existence of several differences statistically incompatible with zero, demonstrating that \gls{mlcb} robustly measures asymmetries between Pauli eigenvalues inaccessible to conventional characterization.

Our experimental results (detailed in App.~\ref{app:exp}) show that the products of Pauli eigenvalues measured by \gls{mlcb} on IQM Garnet\textsuperscript{\tiny{TM}} deviate from unity by approximately $0.1$. However, the observed differences $\Delta_q^{L_1L_2}$ are one-to-two orders of magnitude smaller, consistently below $0.006$. This suggests that while \gls{mlcb} provides a statistically significant improvement in accuracy over conventional methods, the full extent of this gain may be partially obscured by other experimental non-idealities, such as temporal instabilities and deviations from the considered effective noise model.
}

\section{Improved error characterization}
\label{sec:impro}
\add{Having theoretically established that \gls{mlcb} can significantly reduce the number of unlearnable \gls{dof} in realistic scenarios and having demonstrated that these protocols can be implemented in practice, we now investigate to what extent this additional knowledge improves the overall accuracy of the fitted \gls{spl} model. This is not a trivial step, as it requires combining high-accuracy data from \gls{cb} and \gls{mlcb} with lower-accuracy methods needed to compensate for the remaining unlearnable components of the Pauli noise.}

\begin{figure}
    \centering
    \includegraphics[width=\linewidth]{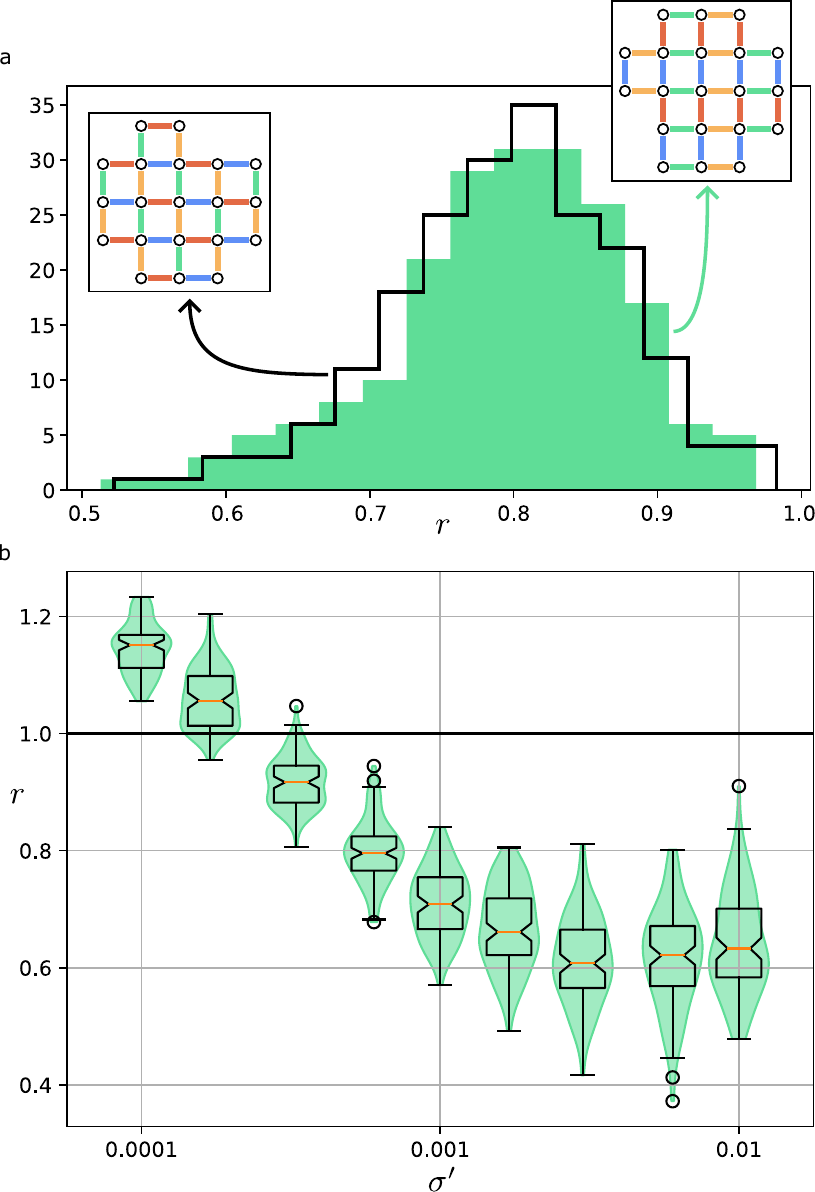}
    \caption{(a) Reduction of the $L_1$ distance between the exact and characterized noise model enabled by \gls{mlcb}\add{, with respect to conventional \gls{cb} approach and symmetry assumption}. We consider $200$ randomly generated noise models and plot the ratio $r = \Delta_m/\Delta_c$ \add{(see Eq.\eqref{eq:Delta})} for each of the two layer configurations shown in the insets. (b) Similarly to the previous panel, we plot ratios $r$ associated with conventional strategies that allow \add{measuring} unlearnable quantities with accuracy $\sigma'\geq \sigma = 10^{-4}$ \add{(e.g. via unit-depth circuits)}, for the configuration shown in left inset of panel (a).}
    \label{fig:better_char}
\end{figure}

We address this point through numerical simulations, which provide a controlled environment free from other experimental non-idealities and where the exact noise model is known for reference—a condition not met in a real device. To this end, we numerically simulate the characterization of a $20$-qubit IQM Garnet\textsuperscript{\tiny{TM}} QPU \cite{Abdurakhimov2024}. In particular, we consider two different dense layer structures, depicted in the insets of Fig.~\ref{fig:better_char}(a). The leftmost configuration corresponds to the one used for the experiment in Sec.~\ref{sec:exp}, with pairs of layers defining open chains, whose \gls{mlcb} analysis follows the scheme in Fig.~\ref{fig:chains}(b). In contrast, the rightmost configuration features closed chains, which are analyzed according to the scheme in Fig.~\ref{fig:chains}(c).

Our numerical simulations consist of five steps. At first, (i) we randomly generate a realistic \gls{spl} noise model, i.e. we associate a map $\Lambda_L$ to each one of the four layer $L\in\{B,G,R,O\}$ based on model parameters $\vec{\lambda}^L$ that mimics the behavior of realistic devices, see Appendix \ref{app:noise} for more details. (ii) From each {exact} noise map $\Lambda_L$ we compute all the (products of) eigenvalues that are needed to fit the \gls{spl} model. In an actual experiment, those would be obtained using conventional \gls{cb}, \gls{mlcb}, as well as possible alternative low-accuracy methods. (iii) We artificially add uncertainties independently to all (product of) eigenvalues, in the form of Gaussian noise, to simulate the statistical uncertainty associated with an actual experimental characterization. For (products of) eigenvalues determined via conventional \gls{cb} and \gls{mlcb}, we consider a fixed standard deviation $\sigma$. To simulate the implementation of low-accuracy strategies that involve performing extra measurements, like the one based on unit-depth \gls{cb}-like circuits, we account for the additional uncertainties due to \gls{spam} noise and the lack of multiplicative precision by considering a larger standard deviation $\sigma'\geq\sigma$. (iv) We then use these noisy eigenvalues to fit two \gls{spl} models for each layer: $\Lambda_L^c$, which is based on parameters $\{\vec{\lambda}^L_{c}\}$ fitted using solely conventional \gls{cb} data and low-accuracy strategies, and $\Lambda_L^m$, whose fitted parameters $\{\vec{\lambda}^L_{m}\}$ also leverages the additional knowledge associated with \gls{mlcb} experiments. (v) Finally, we compute the $L_1$ distances $\Delta_t$, with $t=c,m$, between the exact and the reconstructed noise models, summed over all the layers $L\in\{B,G,R,O\}$ as
\begin{equation}
\label{eq:Delta}
\Delta_t = \sum_{L} ||\vec \lambda^L - \vec{\lambda}^L_t||_1.
\end{equation}
We also define their ratio as $r = \Delta_{m}/\Delta_{c}$.

The most intricate aspect of this otherwise straightforward procedure lies in step (iv), particularly the role of products of eigenvalues measured with \gls{mlcb} during the model fitting process. Several strategies can be adopted to address this step. One straightforward method is to treat experimental data from conventional \gls{cb}, \gls{mlcb}, and low-accuracy methods on the same ground, by generalizing Eq.~\eqref{eq:vec_f<>vec_lambda} to the multi-layer scenario
\begin{equation}
\label{eq:gen_xi_lambda}
    \log(\vec{\xi}^+) = -2 \bar{M}^+ \vec{\lambda}^+.
\end{equation}
Here, $\vec{\lambda}^+$ is the concatenation of the parameter vectors ${\vec\lambda}_L$, associated with each individual layer $L\in\{B,G,R,O\}$. Similarly, $\vec{\xi}^+$ is the concatenation of the (products of) eigenvalues vectors $\vec{\xi}_L$, which are used in the conventional approach, supplemented with the products of eigenvalues $\mathcal{F}_\alpha^{L_1L_2}$ measured using \gls{mlcb}. Eq.~\ref{eq:gen_xi_lambda} can be then solved using non-negative least square fitting in a $4|\mathcal{K}|$-dimensional space.

Alternatively, for each layer $L$, one can first identify a specific set of unlearnable eigenvalues $\{f^L_\beta\}$ required to fit the \gls{spl} model. For example, as discussed above, one can focus on all weight-one Pauli strings $\beta$ consisting of a single $X$ operator on the layer's support. Since these eigenvalues are inherently unlearnable, their initial estimates rely on low-accuracy data. However, they can be significantly improved by imposing the high-accuracy constraints provided by \gls{mlcb}, e.g., $\mu_q^{(L_iL_j)}$. This optimization step, detailed in Appendix \ref{app:fit}, yields refined eigenvalue estimates $\{f^L_\beta\}$, which can then be used for fitting the \gls{spl} model layer-by-layer. In the remainder of this work, we primarily adopt this second approach, which places greater emphasis on leveraging high-accuracy \gls{mlcb} data over low-accuracy estimates. However, other fitting strategies are certainly possible. A thorough exploration of such strategies represents a compelling avenue for future research, as it could unlock further improvements in model fitting accuracy and performance.

The numerical results are summarized in Fig.\ref{fig:better_char}, where we plot the values of the ratio $r$ for different scenarios. In panel (a) we visualize the values of $r$ for $200$ randomly generated noise model, considering both the layer structures shown in the two insets. We consider a reasonable experimental uncertainty of $\sigma = 10^{-4}$ \cite{vandenBerg2022Mar} and we use the symmetry assumption to estimate the unlearnable individual eigenvalues. Regardless of the specific layer structure, we observe a $r$ consistently below $1$, showing that the possibility to measure with high accuracy extra \gls{dof}, unlocked by \gls{mlcb}, leads to a better overall error characterization. 
In Fig.~\ref{fig:better_char}(b), we simulate a characterization procedure that \add{relies} on extra measurements, for example via unit-depth \gls{cb}-like experiments, to estimate with low-accuracy the required unlearnable eigenvalues. For a fixed $\sigma = 10^{-4}$, we consider different uncertainties associated with unlearnable quantities $\sigma\leq \sigma'\leq 100\sigma $ and compute the ratio $r$ for $100$ randomly generated noise models for each value of $\sigma'$. Provided that $\sigma'\gtrsim 3 \sigma$, which makes the distinction between high- and low-accuracy procedures meaningful and justified, we observe values of the ratio $r$ well below $1$, which saturates to a range $0.5 \lesssim 0.8$ in the realistic regime $\sigma'\gtrsim 10 \sigma$. This shows again how \gls{mlcb} can significantly improve the error characterization. 
By contrast, in the extreme and {unrealistic} regime $\sigma \sim \sigma'$, \gls{mlcb} can even be disadvantageous due to the amplification of the $\sigma$ error associated with the computation of ratios $\mu_1^{(L_1,L_j)}$ (see App.~\ref{app:equivalence}) that makes it more convenient to just consider the direct measurements of the unlearnable eigenvalues. 

\section{Effects on error mitigation}
\label{sec:em}
Error characterization plays a crucial role in informing noise-aware error mitigation strategies, prompting an important question: how does \gls{mlcb} and the associated improved characterization impact error mitigation performance? To address this, we conducted a series of numerical experiments analyzing the performance of \gls{pec}~\cite{Temme2017Nov, vandenBerg2022Mar, Kim2023, Ferracin2022} on Clifford circuits. The choice of Clifford circuits is particularly advantageous because evaluating \gls{pec} performance in this context is both conceptually straightforward and computationally efficient \cite{Govia2024}. Moreover, generic non-Clifford circuits can always be decomposed into a sum of maps associated with Clifford circuits, making the following analysis relevant, at least at a qualitative level, also in practical and general use cases. More precisely, data obtained on Clifford circuits can be indicative of worst-case scenarios performances \cite{Govia2024}.

In the analysis we focus on circuits consisting of the four (dressed) Clifford layers depicted in the rightmost inset of Fig.~\ref{fig:better_char}(a), each one associated with a randomly generated noise map $\Lambda_L$. As discussed in the previous section, for each layer $L\in\{B,G,Y,R\}$ we also define a low-accuracy reconstructed noise map $\Lambda_L^c$ as well as another reconstructed map $\Lambda_C^m$ that makes use of the extra high-accuracy $\gls{dof}$ provided by \gls{mlcb}. When using \gls{pec} to mitigate errors on a given layer, we probabilistically modify the circuit such that the noisy execution of the layer is effectively composed with the inverse of the characterized noise map $\Lambda_L^{t}$, with $t\in\{c,m\}$. As a result, the execution of each layer is effectively associated with the map $ \xi_L^{(t)} =\Phi(U_L)\circ\Lambda_L\circ(\Lambda_L^{t})^{-1}$, which corresponds to the noiseless layer execution in the limit of perfect characterization, i.e. $\Lambda_L^{t} = \Lambda_L$. The effect of such a map on a Pauli operator $P_\alpha$ simply reads
\begin{equation}
\xi_L^{(t)}[P_\alpha] = P_{\alpha'} R_{\alpha, t}^{L} 
\end{equation}
where $P_{\alpha'} = \Phi(U_c)[P_\alpha]$ is the conjugation of $P_\alpha$ under the Clifford unitary and 
\begin{equation}
    R_{\alpha, t}^{L}  = \frac{f_{\alpha}^L}{f_{\alpha, t}^L}
\end{equation} is the ratio between the exact Pauli eigenvalue $f^L_\alpha$, derived from $\Lambda_L$, and the eigenvalue $f_{\alpha, t}^L$ obtained from the characterized model $\Lambda_L^{t}$, with $t\in\{c,m\}$.

We consider Clifford circuits, consisting of the repeated implementations of $J$ Clifford layers $\bar L_j$, with   $j\in[1,J]$. Each $\bar L_i$ is the concatenation of a layer randomly sampled from $\{B,G,R,O\}$ with a random layer consisting of \add{single-qubit} Clifford gates. By preparing a $+1$ eigenstate of a Pauli operator $P_{\beta^{(0)}}$ and measuring the expectation value $O$ of its conjugation $P_{\beta^{(J)}}$ with respect to all Clifford layers, we obtain (neglecting \gls{spam} errors and applying \gls{pec})
\begin{equation}
O_t = \prod_{i=0}^{J-1} R_{\beta^{(i)}, t}^{L}.
\end{equation}
Here $\beta^{(i)}$ is the Pauli string obtained under conjugation by the first $i\geq0$ Clifford layers, i.e. $P_{\beta^{(i)}} = \Phi(\prod_{j=1}^i U_{\bar L_j})[P_{\beta^{(0)}}]$. A perfect characterization would result in the noiseless result $O=1$ and deviations from this value are a consequence of the finite accuracy of the reconstructed maps $\Lambda_L^t$. 

\begin{figure}
    \centering
    \includegraphics[width=1\linewidth]{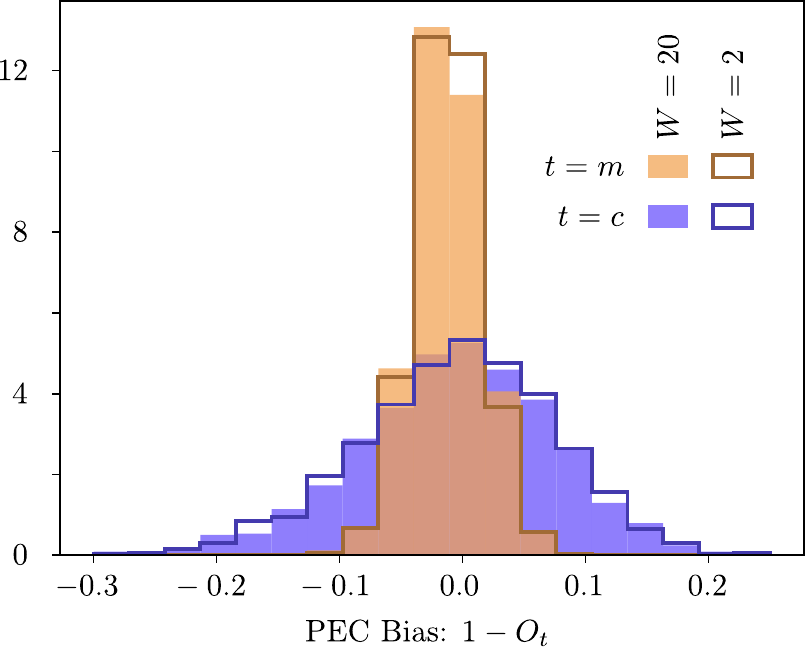}
    \caption{Bias $1-O_t$ resulting from the application of \gls{pec} on random Clifford circuits under $200$ random \gls{spl} noise models, comparing two characterization methods: the conventional procedure ($t=c$) and the advanced one leveraging the additional learnability provided by \gls{mlcb} ($t=m$). The \gls{mlcb}-based method achieves improved error mitigation accuracy, consistently outperforming the standard procedure regardless of the observable's weight, whether low ($W=2$) or high ($W=20$).}
    \label{fig:pec}
\end{figure}

By computing $O_c$ and $O_m$ for different noise models and different circuits, we observed that the latter features a much smaller standard deviation, indicating that the extra learnability offered by \gls{mlcb} can directly result into a better performance of \gls{pec}. The results of this numerical analysis are shown in Fig.~\ref{fig:pec}, where we consider the same $200$ randomly generated noise models employed for Fig.~\ref{fig:better_char}(a). For each noise model, we calculated $O_t$ across ten random circuits, with each circuit composed of 40 random Clifford layers $\bar L_j$. Additionally, each circuit began with a randomly chosen initial Pauli string $\beta^{(0)}$ , subject only to the condition that the final measured Pauli string $\beta^{(J)}$ had a specified weight $W$. The latter parameter does not seem to play a significant role. The standard deviation associated with $O_c$ is $\text{std}(O_c) = 0.08$ for both $W=2$ and $W=20$. By contrast, the standard deviation associated with $O_m$ is $\text{std}(O_m) = 0.03$ for both $W=2$ and $W=20$, showing almost a three-fold improvement over the conventional scenario.

\section{Discussion}

In this paper, we have introduced the \gls{mlcb} protocol and showed its potential in increasing the accuracy of characterization of effective Pauli noise models, thanks to improved learnability. Numerical simulations and a simple experiment have been performed focusing on practically relevant scenarios, featuring dense Clifford layers consisting of parallel \cz gates implemented on a \gls{qpu} with square topology. In these cases, we show in detail how to set up the \gls{mlcb} procedure to significantly decrease the number of unlearnable \gls{dof} while keeping the runtime overhead at a minimum. Our results showcase a scalable and resource-efficient procedure that can be readily extended to various combinations of \cz layers {with minimal overhead}.

While our focus has been on \cz gates and (\gls{spl}) models with weight-two generators, \gls{mlcb} can be generalized to other Clifford gates, higher-weight \gls{spl} models or alternative effective Pauli noise frameworks like CER \cite{CarignanDugas2023}. These extensions, though valuable, are beyond the scope of this paper. On the other hand, \gls{mlcb} is naturally compatible with generalizations of \gls{pt} beyond the Pauli group {with which} the independent number of noise parameters {can be reduced}~\cite{Berg2023}, as exemplified in Section~\ref{sec:spl} for \cz gates and considering the addition of $S$ gates to the twirling set.

It is also important to recognize that limited learnability is not the sole factor affecting characterization accuracy and the performance of noise-aware error mitigation: time-dependent parameter drift, as observed in our experiment, and out-of-model noise contributions can also play a significant role. Recent work, such as Ref.~\cite{Govia2024}, has explored these two issues in the context of \gls{pec}. In this sense, our discussion in Section~\ref{sec:impro} complements these analyses by focusing on the effect of limited learnability. In general, it is important to stress that practical implementations must consider all these factors holistically to ensure robust characterization and effective noise mitigation.

The \gls{mlcb} implementation presented in this work is both practical and resource-efficient, but there is room for optimization in several areas.

First, we have not addressed the optimal allocation of experimental resources, such as the distribution of shots and rounds of \gls{rc}, between different \gls{cb} and \gls{mlcb} protocols. In this regard, insights from the statistical analyses in protocols like \gls{aces}, as detailed in \cite{Hockings2024}, provide valuable directions for future work. Clearly, optimizing the resource allocation could enhance the precision of characterization within a finite experimental runtime. \add{The implementation of \gls{pt} at the hardware level, reaching the ideal limit of a single shot per randomized circuit, is also likely to significantly boost the precision of the characterization protocol with a fixed runtime\cite{Fruitwala2024Jun, Granade2015Jan}}. More broadly, the study of sampling complexity associated with quantum noise characterization is an active and compelling area of research, with recent advancements detailed, e.g., in \cite{Chen2022Mar, Chen2023, Chen2023a}.

Second, as discussed in Section~\ref{sec:impro}, efficiently fitting noise model parameters based on measured (products of) eigenvalues is a non-trivial task. While we propose a weighting strategy that prioritizes high-accuracy data (obtained through \gls{cb} and \gls{mlcb}) over lower-accuracy data from other methods, this approach, validated through numerical simulations of random noise models, is not guaranteed to be optimal. Further exploration in this area could yield more robust and efficient fitting techniques, enhancing the overall utility of \gls{mlcb}.

Finally, while our implementation significantly reduces the number of unlearnable eigenvalues with only a minor runtime increase, we do not exclude the possibility of alternative strategies that could unlock even more \gls{dof}, potentially at the cost of increased experimental runtime. In this respect, during the finalization of this work, we became aware of theoretical advancements in \cite{Chen2024Oct}, which propose a general framework for computing the learnability of Pauli noise models, considering also the \gls{spam} parameters as target of learning. Our results are consistent with their findings in overlapping cases, such as the characterization of quasi-local noise on $n$-qubit systems arranged in 1D rings with two dense layers of parallel \cz gates, thus suggesting the optimality of \gls{mlcb} in this scenario. In general, we believe our work complements the theoretical framework in \cite{Chen2024Oct} by providing practical details for characterizing general layers on 2D \gls{qpu}s with square topologies, thus enabling straightforward implementation on physical devices, and by quantifying the impact of \gls{mlcb}'s enhanced learnability on characterization quality and its effectiveness in error mitigation within the widely adopted \gls{spl} model.

\section{Acknowledgment}
A.C. gratefully acknowledges S. Chen and A. Kandala for their insightful discussions during the APS March Meeting 2024, where the foundational ideas of this work have been presented. The authors acknowledge support from the German Federal Ministry of Education and Research (BMBF) under Q-Exa (grant No. 13N16062) and QSolid (grant No. 13N16161). The authors also acknowledge the entire IQM Technology team for their support in the development of this work.

\appendix
\bibliography{ref}

\glsresetall

\section{Pauli twirling}
\label{app:twirling}
Here we briefly detail how the noise associated with the execution Clifford layer $C$ can be recast into a Pauli channel via twirling. Formally, the twirling of the map $\phi(U_C)$ is performed by sandwiching $C$ in between two layers of \add{single-qubit} operators randomly sampled from the twirling set $\mathbb{T}$. Unless explicitly stated otherwise, in this work we focus on the standard \gls{pt}, with $\mathbb{T}$ consisting of all possible product combination of \add{single-qubit} Pauli {operators}, i.e. $\mathbb{T} =\mathbb{P}_n := \{I,X,Y,Z\}^{\otimes n}$. Instead of executing $U_C$, we first implement a twirling layer $T\in \mathbb{T}$, followed by $U_C$ and then by the conjugated layer $T_C := U_C T U_C^\dagger\in\mathbb{T}$, which also belongs to $\mathbb{T}$ since the Pauli group is invariant under Clifford conjugation. If we assume that the noise map $\Omega$ associated with the execution of a layer $T$ of \add{single-qubit} gates is gate-independent, i.e. $\nu(T) := \Omega\circ\phi(T)$ $\forall T$, one can show that {uniformly sampling $T$ from the twirling set $\mathbb{T}$ and subsequently averaging the results, denoted as 
$\langle\cdot\rangle_{T\in\mathbb{T}}$,} leads to the following effective evolution 
\begin{equation}
\begin{split}
    &\langle \nu(T_C)\circ \mu(U_C)\circ \nu(T) \rangle_{T\in\mathbb{T}} \\
    &\qquad =\Omega\circ \phi(U_C)\circ \langle \phi(T)\circ \tilde\Lambda_C  \circ \Omega\circ \phi(T) \rangle_{T\in\mathbb{T}}  \\
    &\qquad =\Omega\circ \phi(U_C)\circ \Lambda_C,  
\end{split}
\end{equation}
where
\begin{equation}
\label{eq:Lambda_C_app}
    \Lambda_C[\rho] = \sum_{\alpha} p_\alpha P_\alpha \rho P_\alpha^\dagger \quad(P_\alpha \in \mathbb{P}_n),
\end{equation}
is an effective Pauli channel associated with the execution of the {Pauli-dressed} Clifford layer $C$. {The channel $\Lambda_C$ corresponds to the noise of the original} Clifford layer $C$, which contributes the majority of the noise, {together with the noise of the twirling layer $T$}.

\section{The example of a single \cz gate}
\label{app:ex_CZ}
\begin{figure}
    \centering
    \includegraphics[width=1\linewidth]{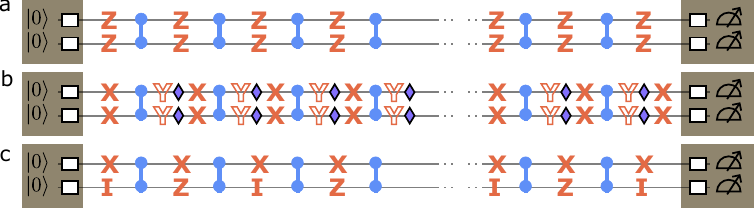}
    \caption{(a) Simple example of a \gls{cb} circuit that allows for the high-accuracy determination of Pauli eigenvalue $f_{ZZ}$. For the sake of clarity, here we do not display the white boxes associated with \gls{rc} and, instead, we show {(}in red) the Pauli operator which is measured at the end and how it propagates back (considering its conjugation under the \cz gates) all the way to the state preparation stage. (b) Similarly to the previous panel, we show how interleaved \gls{cb} can determine the eigenvalue $f_{XX}$ with high accuracy. The extra purple diamonds show $S$ gates that are applied after every \cz (and eventually recompiled together with the non-displayed layer of \add{single-qubit} gates) which map the Pauli operators $P_{YY}$ and $P_{XX}$ into each other. (c) In this case, \gls{cb} protocols can only determine with high accuracy the product of eigenvalues $f_{XI}f_{XZ}$, which cannot be individually measured with high accuracy.}
    \label{fig:cb_cz}
\end{figure}

Let us provide more detail about the simple-yet-instructive example of a system with $n=2$ qubits and on a layer $C$ consisting of a single \cz gate between the two qubits. The fifteen non-trivial Pauli operators $P_\alpha \in \mathbb{P}_2$ are listed in the first row of Tab.~\ref{tab:conjugation} in the main text, while the second row shows their conjugation under $C$. Since there are only three orbits with cardinality one, associated with the Pauli strings $\alpha \in \{IZ,ZI,ZZ\}$, the three corresponding Pauli eigenvalues are the only ones that can be determined with high accuracy using standard \gls{cb} directly on the layer $C$. See, for example, the setup in Fig.~\ref{fig:cb_cz}(a) that can measure $A_{ZZ}(f_{ZZ})^d$ and, thus, determine with high accuracy the individual eigenvalue $f_{ZZ}$. All other orbits have cardinality $2$ (which is the largest possible since, in this case, $U_C^2=I$). In several cases, however, the Pauli operators belonging to the same orbit share the same \add{two-qubit} support (for example, $\mathcal{R}^C_{XX} = \mathcal{R}^C_{YY} = \{P_{XX},P_{YY}\}$). In these cases, the orbit's cardinality can be reduced to one by inserting two extra $S$ gates after each execution of the layer $C$ (and then recompiling them with the following layer of \add{\add{single-qubit}} gates). This modifies the associated unitary as $U_{\bar C} = (S\otimes S) U_C$, providing us with the third row of Tab.~\ref{tab:conjugation} and, importantly, with four novel orbits with size one: $\mathcal{R}^{\bar C}_\alpha$ for $\alpha \in \{XX,XY,YX,YY\}$. An example of an interleaved \gls{cb} circuit that allows determining $f_{XX}$ is shown in Fig.~\ref{fig:cb_cz}(b). At this point, we are left with eight Pauli strings, which {can be} conveniently organized into two sets $\mathcal{S}_A = \{IX,IY,ZX,ZY\}$ and $\mathcal{S}_B = \{XI,YI,XZ,YZ\}$, whose associated orbits cannot be reduced to size one. This means that the Pauli eigenvalues corresponding to the elements of each set therefore cannot be determined individually with high accuracy. However, it is important to stress that, out of the $8$ DOF associated with these $8$ eigenvalues, the \gls{cb} protocols still allows us to determine $6$ independent high-accuracy constraints on the products of eigenvalue pairs, thus leaving us with only $2$ unlearnable DOF. Indeed, focusing on set $\mathcal{S}_A$, one can determine the product $f_{IX}f_{ZX}$, as shown in Fig.~\ref{fig:cb_cz}(c), as well as $f_{IY}f_{ZY}$ and $f_{IX}f_{ZY}$ (the fourth product is not independent since $f_{IY}f_{ZX} = (f_{IY}f_{ZY})(f_{IX}f_{ZX})/(f_{IX}f_{ZY})$). The same applies to the other set $\mathcal{S}_B$. All in all, this example shows that a single \cz gate is associated with $2$ unlearnable DOF, which cannot be determined using CB protocols and which can be (non-unambiguously) identified with eigenvalues $f_{IX}$ and $f_{XI}$. These results mirror the findings discussed in Ref.~\cite{Chen2023Jan} for a single CNOT gate. 

\section{Equivalence between products of eigenvalues}
\label{app:equivalence}

The objective of this section is to establish the equivalence between products of eigenvalues directly measurable via \gls{mlcb}---such as  $o_3, o_4, o'_4, o_5, c_4$ introduced in the main text---and specific ratios of unlearnable eigenvalues, $\mu_q^{(BG)}$. As established at the end of Sec.~\ref{sec:simple_ex}, two functions of eigenvalues are deemed equivalent if one can be fully derived from the other using only high-accuracy data, i.e. (products of) learnable eigenvalues.

To achieve this, we first introduce an equivalence test to formally prove the equivalence between functions of eigenvalues within a single layer. We then extend this test to multi-layer scenarios and demonstrate its application using a concrete example of three-qubit chains. This example illustrates each step required to prove equivalence and establish relationships such as those expressed in Eq.\eqref{eq:equiv_2}. Building on this example, we generalize the methodology to more complex cases and summarize the results.

\subsection{Equivalence test}
We begin by focusing on a single layer $C$ and linear combinations of logarithms of eigenvalues. These combinations can be written as $\zeta_Y = \sum_{\alpha \in \mathcal{X}_Y} \gamma_\alpha \log f^C_\alpha$, where $\gamma^Y_\alpha$ are numerical coefficients and $\mathcal{X}_Y$ is a set of Pauli strings. 

Within the \gls{spl} model, the vector of model parameters $\vec \lambda$ determines all functions of eigenvalues via Eq.~\eqref{eq:vec_f<>vec_lambda}. In particular, the vector $\vec \Xi$, consisting of all learnable (products of) eigenvalues which can be determined, for instance, by following the strategy outlined in Section \ref{sec:spl}, is determined by 
\begin{equation}
    \log(\vec \Xi) = \bar M \vec \lambda.
\end{equation}
Analogously, $\zeta_Y$ can be determined from the model parameters via
\begin{equation}
    \zeta_Y = (\vec m_Y)^T \vec \lambda
\end{equation}
where the $\kappa$-th element of the single-row matrix $(\vec m_Y)^T$ reads 
\begin{equation}
    (\vec m_Y)_\kappa = \sum_{\alpha \in \mathcal{X}_Y} \gamma_\alpha \langle \alpha, \kappa \rangle_{sp}.
\end{equation}

To test whether two functions of eigenvalues, $\zeta_A$ and $\zeta_B$, are equivalent, we compute $\bar M$, $\vec m_A$ and $\vec m_B$. Let $r$ be the rank of $\bar M$; $r_Y$ the rank of matrix $M$ concatenated with the extra row $(\vec m_Y)^T $; $r_{AB}$ the rank of the matrix $\bar M$ concatenated with $(\vec m_A)^T $ and $(\vec m_B)^T $. The equivalence condition is then
\begin{equation}
\label{eq:equivalence_condition}
\zeta_A \text{ and } \zeta_B \text{ are equivalent} \Leftrightarrow r_A=r_B=r_{AB}.
\end{equation}
Indeed, $r_A = r_{AB}$ shows that $(\vec m_B)^T$ is a linear combination of the rows of $\bar M$ and of $(\vec m_A)^T$, meaning that $\zeta_B$ can be expressed in terms of $\zeta_A$ and some learnable quantities in $\vec \Xi$. The same logic applies symmetrically for $A$ and $B$.

Notably, if $r_Y = r$, then $\zeta_Y$ is fully learnable and provides no new information for fitting the model parameters. 

\begin{figure}
    \centering
    \includegraphics[width=\linewidth]{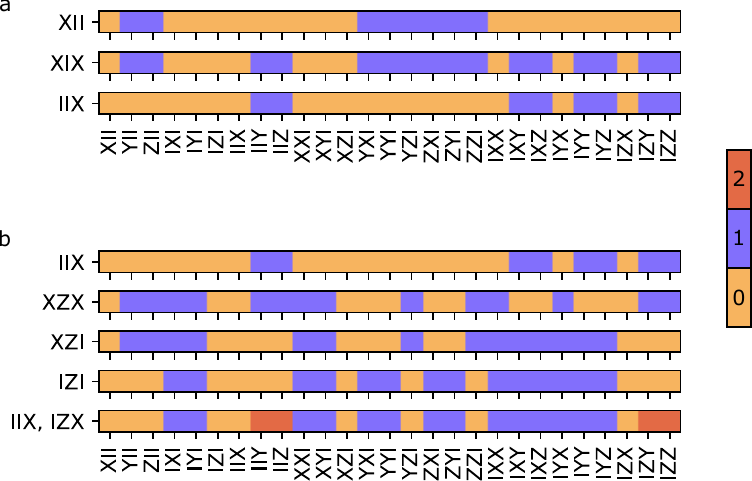}
    \caption{Visualization of single-row matrices $(\vec m_{\alpha})^T$ associated with the blue layer (panel a) and the green layer (panel b) for the three-qubit open chain depicted in Fig.~\ref{fig:MLCB}(c). On the horizontal axes, all the $27$ ordered Pauli strings $\kappa \in \mathcal{K}$ are listed. The first nine consists are weight-one Pauli strings, the second nine are weight-two Pauli strings with support on qubits $0$ and $1$, and the last nine are weight-two Pauli strings with support on qubits $1$ and $2$. As for the Pauli strings $\alpha$, they are specified to the left of each row. Note that the last row actually depicts $(\vec m_j^T) = (\vec m_{IIX}^T) + (\vec m_{IZX}^T)$.}
    \label{fig:Mat}
\end{figure}

\subsection{Three-qubit open chain}
We now consider the three-qubit open chain described in Sec.~\ref{sec:simple_ex}. Our goals are: (i) to prove the equivalence between $o_3 = f^B_{XIX}f^G_{XZX}$ and $\mu_1^{(BG)}=f^B_{XII}/f^G_{IIX}$ (ii) to prove Eq.~\eqref{eq:equiv_2} which directly links these quantities. 

For concreteness, we assume a \gls{qpu} with square topology, ensuring qubits $0$ and $2$ are not neighbours. In this case, the \gls{spl} model describing the noise in the chain features a total of $|\mathcal{K}| = 27$ parameters, associated with the Pauli strings $\kappa$ listed on the $x$-axes of the plots in Fig.~\ref{fig:Mat}. 

We first analyze the blue layer, focusing on the function of (now trivial) eigenvalues $F^B_1 = f^B_{XII}$ and $F^B_2 = f^B_{XIX}$. Computing the corresponding matrix rows 
\begin{align}
    (\vec m_{XII}^T)_\kappa = \langle XII, \kappa\rangle_{sp}\\
    (\vec m_{XIX}^T)_\kappa = \langle XIX, \kappa\rangle_{sp}
\end{align}
we verify the equivalence condition from Eq.~\eqref{eq:equivalence_condition}. This proves $\log(F^B_1)$ can be expressed as a linear combination of $\log(F^B_2)$ and other function of learnable eigenvalues. To explicitly identify the latter, we implement a simple search algorithm, which progressively attempts to remove (randomly chosen) rows of matrix $\bar M$ while still preserving the condition on the ranks in Eq.~\eqref{eq:equivalence_condition}. We find that, in this case, the single row of matrix $\bar M$ associated with the learnable eigenvalue $f^B_{IIX}$, i.e. 
\begin{equation}
    (\vec m_{IIX}^T)_\kappa = \langle IIX, \kappa\rangle_{sp},
\end{equation}
is sufficient to relate $(\vec m_{XII}^T)$ and $(\vec m_{XIX}^T)$ via $(\vec m_{XII}^T) = (\vec m_{XIX}^T) - (\vec m_{IIX}^T)$. This is visualized in Fig.~\ref{fig:Mat}(a). In view of the subsequent analysis, it is convenient to summarize this result as 
\begin{equation}
\label{eq:F_L}
    \log(F^L_1) = \epsilon^L \log(F^L_2)
    + (\vec{\Sigma}^L)^T \log(\vec{F}^L)
\end{equation}
with $L=B$, $\epsilon^B = 1$, $\vec{\Sigma}^B = (-1)$ and $\vec{F}^B = (f^B_{IIX})$. This proves \begin{equation}
f^B_{XII} = \frac{f^B_{XIX}}{f^B_{IIX}}
\end{equation}

Similarly, for the green layer we focus on functions $F^G_1 = f^G_{IIX}$ and $F^G_2 = f^G_{XZX}$, proven to be equivalent by \eqref{eq:equivalence_condition}. Moreover, we show that the corresponding matrix rows $(\vec m_{IIX}^T)$ and $(\vec m_{XZX}^T)$ can be related to each other by using three rows of matrix $\bar M$. Specifically, we obtained 
\begin{equation}
    (\vec m_{IIX}^T) = - (\vec m_{XZX}^T) + (\vec m_{XZI}^T) -  (\vec m_{IZI}^T)  + (\vec m_j^T) .
\end{equation}
where the last row $(\vec m_j^T) = (\vec m_{IIX}^T) + (\vec m_{IZX}^T)$ is associated to the learnable {product} of eigenvalues in the orbit $\mathcal{R}^G_{IIX} = \{P_{IIX}, P_{IZX}\}$. Such a relation is visualized in Fig.~\ref{fig:Mat}(b). This result can be summarized in the form of Eq.~\eqref{eq:F_L}, with $L=G$, $\epsilon^G = -1$, $\vec{\Sigma}^G = (1,-1,1)$ and $\vec{F}^G = (f^G_{XZI}, f^G_{IZI}, f^G_{IIX} f^G_{IZX})$. This proves 
\begin{equation}
f^G_{IIX} = \frac{f^G_{XZI}(f^G_{IZX}f^G_{IIX})}{f^G_{XZX}f^G_{IZI}}
\end{equation}

By combining these results, valid for each layer individually, we obtain Eq.~\ref{eq:equiv_2}, thus showing the equivalence between $o_3 = F^B_2 F^G_2$ and $\mu_1^{(BG)} = F^B_1/F^G_1 $. It is important to stress that the equivalence class does not consists only of these two functions of eigenvalues. It is indeed possible to find other equivalent functions of eigenvalues, all associated to the same \gls{dof}, whose learnability is unlocked by \gls{mlcb}. For example, one has
\begin{equation}
   f^B_{XII} f^G_{IZX} = \mu_1^{(BG)} (f^G_{IIX} f^G_{IZX}),
\end{equation}
where $f^G_{IIX} f^G_{IZX}$ is learnable with conventional \gls{cb}.

\subsection{Other chains}
Longer chains introduce additional challenges due to the connectivity of the underlying \gls{qpu}. Non-neighboring qubits in the chain may still interact via the hardware's topology, increasing the number of \gls{spl} parameters and altering the relationships between functions of eigenvalues. Relevant examples are shown in the first column of Tables \ref{tab:4qo}-\ref{tab:4qc}

In the following, we analyze all relevant chain configurations, considering a \gls{qpu} with square topology. For each case, we apply the methodology outlined for the three-qubit chain. Specifically, focusing on one layer at a time, we use the equivalence test to demonstrate the equivalence between two functions of eigenvalues, denoted as $F^L_1$ and $F^L_2$, respectively. 
This approach enables us to prove the equivalence—and derive a direct connection—between the eigenvalue product measurable via \gls{mlcb}, $F^B_2 F^G_2$, and the ratios of unlearnable eigenvalues, $F^B_1/ F^G_1$. The results for all chain configurations are summarized in Tables \ref{tab:4qo}-\ref{tab:4qc}.

It is important to note that the reported $\vec\Sigma^L$ and $\vec{F}^L$ are not unique, as alternative choices for these quantities can still satisfy Eq.~\eqref{eq:F_L}. An interesting open question concerns the optimization of the choice of $\vec\Sigma^L$ and $\vec{F}^L$, with the goal of minimizing the number of components in each vector for a given chain. While this optimization does not affect the learnability of eigenvalues and is thus beyond the scope of this paper, it can be beneficial when implementing the characterization procedure in practice. For instance, in the four-qubit closed chain analyzed in Table~\ref{tab:4qc}, connecting $c_4 = f^B_{IXXX} f^G_{ZXYY} f^B_{IYYX} f^G_{ZYXY}$ to the simple eigenvalue ratio $\mu_q^{(BG)}$ requires considering a total of $28$ learnable (products of) eigenvalues. Reducing this number could streamline the analyses and make them more robust with respect to the spread of statistical uncertainties. 

\section{Experimental details}
\label{app:exp}
In this appendix, we provide further details on the experimental execution and data analysis of the \gls{mlcb} protocol discussed in Sec.~\ref{sec:exp}.

For each of the six pairs of layers $(L_1, L_2)$ depicted in Fig.~\ref{fig:exp}(a), we perform a complete MLCB experiment. Each experiment consists of running circuits for five different sequence depths $d \in {2,4, 8, 16, 32}$. For each depth, we use a total of $4 \times 10^4$ measurement shots, which are distributed evenly across 20 random Pauli twirling instances. This results in $2000$ shots per unique circuit configuration. To gather more statistics and develop a feeling of the time stability of the results, we repeat this entire procedure three times back-to-back for every layer pair.

\begin{figure} \centering 
\includegraphics[width=\linewidth]{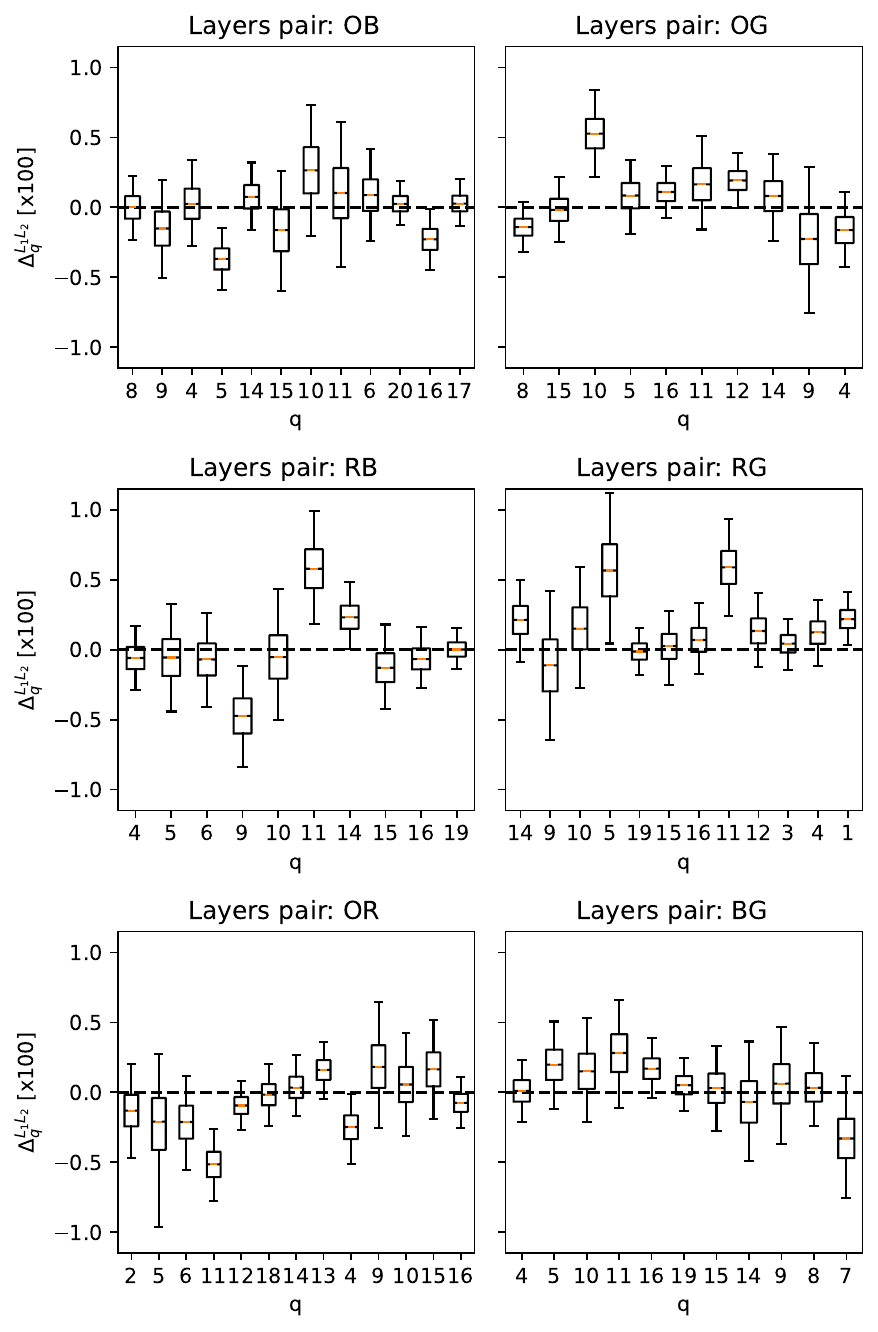} \caption{\add{Summary of all measured asymmetry values $\Delta_q^{L_1L_2}$. Each panel corresponds to a different pair of layers, as indicated by the title. The box plots show the distribution of $\Delta_q$ for each hot qubit $q$ in the chain, averaged over three consecutive MLCB experiments. The whiskers indicate the 95\% confidence interval derived from bootstrapping.}} \label{fig:appendix_deltas} \end{figure}

The statistical uncertainty of our measurements is estimated using a bootstrapping procedure. For each of the three experimental runs, we generate 100 bootstrap samples by resampling with replacement from both the measurement shots and the 20 random twirling instances. The final reported values and confidence intervals for $\Delta_q^{L_1L_2}$ are derived from the combined distribution of these bootstrapped results, averaged across the three consecutive runs. 
In Fig.~\ref{fig:appendix_deltas}, we present the complete set of measured $\Delta_q^{L_1L_2}$ values for all six layer pairs, averaged over the three consecutive experimental runs.

To assess the normality of the bootstrapped distributions, we compute the skewness $s_k$ and excess kurtosis $k_u$ for each difference $\Delta^{L_1L_2}_{q}$. With the exception of $\Delta^{\rm OR}_{5}$, which exhibits large skewness $s_k = -1$ and excess kurtosis $k_u = 2.5$ (as also evident also from the asymmetric whiskers in Fig.~\ref{fig:appendix_deltas}), 
all the other differences features small values within the ranges $-0.18 \leq s_k \leq 0.25$ and $-0.14 \leq k_u \leq 0.32$.
This confirms the approximate normality of the bootstrapped distributions, thereby validating the use of the associated ratios $\Omega^{L_1L_2}_{q}$, between mean and standard deviations, to assess their statistical compatibility with zero. 

\begin{figure} \centering 
\includegraphics[width=\linewidth]{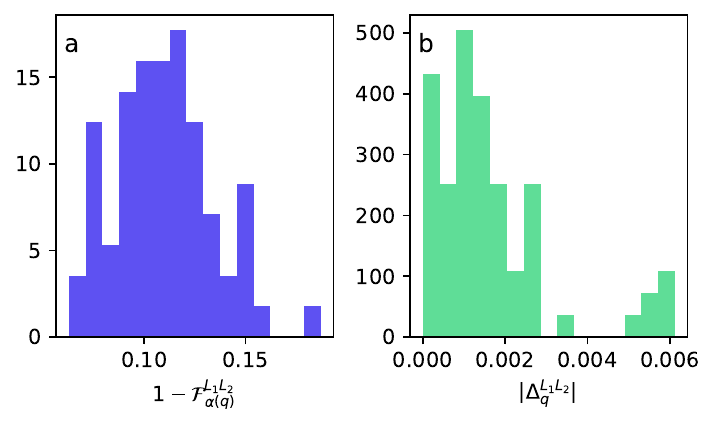} \caption{\add{(a) Distribution of all the ``infidelities" $1-\mathcal{F}_{\alpha(q)}^{L_1L_2}$ (panel a) and of all the absolute differences $|\Delta_{q}^{L_1L_2}|$ (panel b) measured by the \gls{mlcb} protocol, for all $q$ and all layer pairs.
}} \label{fig:appendix_deltas_hist} \end{figure}

As detailed in Fig.~\ref{fig:appendix_deltas_hist}(a), the products of Pauli eigenvalues measured by \gls{mlcb} deviates from $1$ by around $0.1$. By contrast, the absolute differences $|\Delta_q^{L_1L_2}|$ remain below $0.006$ (see Fig.~\ref{fig:appendix_deltas} and Fig.~\ref{fig:appendix_deltas_hist}(b)). While these deviations are statistically significant, as established above, their magnitude indicates that violations of the symmetry assumption between unlearnable quantities contribute only a few percent to the overall value of the Pauli eigenvalues.

Let us now illustrate with a detailed example what is accessible to \gls{mlcb} and standard \gls{cb} protocols in terms of products of Pauli eigenvalues. To this end, we focus on the layers pair BR and on qubit $q=11$, i.e. to the case analyzed in detail in Fig.~\ref{fig:exp}(c). Here we have
\begin{equation}
\mathcal{F}_{\alpha(11)}^{BR} = f_{IXIX}^Bf_{IXZX}^Rf_{ZXIX}^Bf_{ZXZX}^R,
\end{equation}
where the Pauli strings refers to qubits $9$, $10$, $11$ and $12$ (see Fig.~\ref{fig:exp}(a)). As $\beta(11) = IXZX$, we also have
\begin{equation}
\mathcal{F}_{\beta(11)}^{BR} = f_{IXZX}^Bf_{ZXIX}^Rf_{ZXZX}^Bf_{IXIX}^R.
\end{equation}
Those two products can be measured with high-accuracy with \gls{mlcb} but cannot be individually accessed via conventional \gls{cb}. The latter could only be used to measure, by combining the analyses of the two layers, the product $f=\mathcal{F}_{\alpha(11)}^{BR} \mathcal{F}_{\beta(11)}^{BR} $ which reads
\begin{equation}
\begin{split}
    f =& (f_{IXIX}^Bf_{IXZX}^B)(f_{ZXIX}^B f_{ZXZX}^B)\\&\quad (f_{IXIX}^R f_{ZXZX}^R)( f_{IXZX}^R f_{ZXIX}^R)
\end{split}
\end{equation}
and all the products within round brackets are learnable with standard \gls{cb}.

\section{Fitting the noise parameters}
\label{app:fit}
Here we detail how we impose the high-accuracy constraint obtained with \gls{mlcb} on the low-accuracy eigenvalues. Let us focus on a qubit $q$, belonging to the support of $l$ layers $L_1, \dots, L_l$. 
When relying exclusively on conventional \gls{cb} protocols, qubit $q$ is associated with $l$ unlearnable \gls{dof}, which can be identified with the $l$ eigenvalues in $\{f^{L_j}_{\beta_q^j}\}$, where the weight-one Pauli strings $\beta_q^j$ feature a single $X$ operator on the qubit connected to $q$ via the $L_j$ layer. Low-accuracy methods, like the symmetry assumption or unit-depth \gls{cb}-like experiments, provide us with estimates for each one of these eigenvalues, which we indicate with an extra tilde $\tilde f^{L_j}_{\beta_q^j}$ and represent the numerical values used for the conventional model fitting. 

\gls{mlcb} allows us to determine $l-1$ additional high-accuracy constraints, which can be identified, for example, with eigenvalue ratios 
\begin{equation}
    \mu_q^{L_j, L_{j+1}} = \frac{f^{L_j}_{\beta_q^j} }{ f^{L_{j+1}}_{\beta_q^{j+1}} }
\end{equation}
for $j=1, \dots, l-1$. This reduces the number of unlearnable \gls{dof} to one, which we can identify with a specific eigenvalue, say $f_q^{L_1}$. We estimate its numerical value $u$ by minimizing the sum $S$ of the squared residues
\begin{equation}
    S = (\tilde f^{L_j}_{\beta_q^j} - u)^2 + \sum_{j=1}^{l-1} \left(
        \tilde f^{L_j}_{\beta_q^j} - u \prod_{i=1}^j \mu_q^{L_i, L_{i+1}} 
    \right)^2
\end{equation}
computed by comparing the low-accuracy eigenvalue estimations with the effect of the high-accuracy constraints from \gls{mlcb}. This procedure can be readily adjusted to tackle scenarios where different pairs of layers have been studied with \gls{mlcb}.
\section{Random noise models}
\label{app:noise}

The random noise models are generated to simulate the behavior of realistic devices. The procedure involves assigning values to all model parameters for a given layer $L$ distinguishing them based on the following criteria:
\begin{itemize}
    \item Weight ($w$) of the Pauli operator $P_\kappa$, either $w=1$ or $w=2$;
    \item Status ($s$) of the involved qubits, which can be either active ($s=a$) if all qubits in the support of $P_\kappa$ belong to the support of $L$, or inactive ($s=i$) otherwise. 
\end{itemize}
Inactive parameters are drawn from a Gaussian distribution with mean $m_w^i$ and standard deviation $\sigma_w^i$. The active ones are sampled from a Gaussian with mean $m_w^a$ and standard deviation $\sigma_w^a$, where, for every gate in the layer, $m_w^a$ itself is drawn from a Gaussian with mean $\bar m_w^a$  and standard deviation $\bar \sigma_w^a$ . This additional sampling step introduces variability between gates within a single layer, resulting in better and worse-performing ones. 

Finally, note that any negative model parameters resulting from the sampling process are set to zero. The parameter values used in this procedure reads
\begin{equation}
\begin{split}
    m_1^i = 2\,10^{-4} \quad & \sigma_1^i = 8\,10^{-4}\\
    \bar m_1^a = 1\,10^{-3} \quad & \bar \sigma_1^a = 7.5\,10^{-4} \\
     \quad & \sigma_1^a = 1\,10^{-3}
\end{split}
\end{equation}
for $w=1$ and
\begin{equation}
\begin{split}
    m_2^i = 1.5\,10^{-4} \quad & \sigma_2^i = 1\,10^{-3}\\
     \bar m_2^a = 2\,10^{-3} \quad & \bar \sigma_2^a = 1.5\,10^{-3} \\
     \quad & \sigma_2^a = 2\,10^{-3}
\end{split}
\end{equation}
for $w=2$.

\input{tables}

\end{document}

%% file: tables.tex
\begin{table*}
\caption{\label{tab:4qo} Four-qubit open chains (blue-green-blue)}
\begin{ruledtabular}
\begin{tabular}{cccccc}
        \multirow{4}{*}{\includegraphics[width=1.5cm]{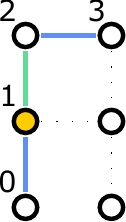}}& $F^B_1$ & $F_2^B$  & $\epsilon^B$ & $\vec{\Sigma}^B$  & $\vec{F}^B$ \\[.15cm]
        & $f^{B}_{XIII}$ & $f^{B}_{XIXI}f^{B}_{XIXZ}$  & $+0.50$ & \(\displaystyle \begin{pmatrix}
-0.50 \\ 
\end{pmatrix} \)  & \(\displaystyle \begin{pmatrix}
f^{B}_{IIXZ}f^{B}_{IIXI} \\ 
\end{pmatrix} \) \\[.15cm]
\cline{2-6} \\[-.25cm]
& $F^G_1$ & $F_2^G$  & $\epsilon^G$ & $\vec{\Sigma}^G$  & $\vec{F}^G$ \\[.15cm]
        & $f^{G}_{IIXI}$ & $f^{G}_{XZXZ}f^{G}_{XZXI}$  & $-0.50$ & \(\displaystyle \begin{pmatrix}
+0.50 \\ 
 -0.50 \\ 
 +0.25 \\ 
 +0.75 \\ 
\end{pmatrix} \)  & \(\displaystyle \begin{pmatrix}
f^{G}_{XZII}f^{G}_{XZII} \\ 
f^{G}_{IZII}f^{G}_{IZII} \\ 
f^{G}_{IIXZ}f^{G}_{IZXZ} \\ 
f^{G}_{IZXI}f^{G}_{IIXI} \\ 
\end{pmatrix} \) \\
\cline{1-6} \\

        \multirow{4}{*}{\includegraphics[width=1.5cm]{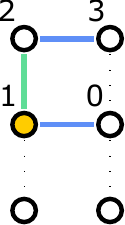}}& $F^B_1$ & $F_2^B$  & $\epsilon^B$ & $\vec{\Sigma}^B$  & $\vec{F}^B$ \\[.15cm]
        & $f^{B}_{XIII}$ & $f^{B}_{XIXI}f^{B}_{XIXZ}$  & $+0.50$ & \(\displaystyle 
        \begin{pmatrix}
+0.25 \\ 
 -0.50 \\ 
 +0.25 \\ 
 -0.25 \\ 
\end{pmatrix}
         \)  & \(\displaystyle \begin{pmatrix}
f^{B}_{IIIZ}f^{B}_{IIIZ} \\ 
f^{B}_{IIXZ}f^{B}_{IIXI} \\ 
f^{B}_{XZII}f^{B}_{XIII} \\ 
f^{B}_{XIIZ}f^{B}_{XZIZ} \\ 
\end{pmatrix} \) \\[.15cm]
\cline{2-6} \\[-.25cm]
& $F^G_1$ & $F_2^G$  & $\epsilon^G$ & $\vec{\Sigma}^G$  & $\vec{F}^G$ \\[.15cm]
        & $f^{G}_{IIXI}$ & $f^{G}_{XZXZ}f^{G}_{XZXI}$  & $-0.50$ & \(\displaystyle \begin{pmatrix}
-0.50 \\ 
 +0.25 \\ 
 +0.75 \\ 
 -0.25 \\ 
 +0.50 \\ 
 -0.25 \\ 
 +0.25 \\ 
\end{pmatrix} \)  & \(\displaystyle \begin{pmatrix}
f^{G}_{IZII}f^{G}_{IZII} \\ 
f^{G}_{IIXZ}f^{G}_{IZXZ} \\ 
f^{G}_{IIXI}f^{G}_{IZXI} \\ 
f^{G}_{IIIZ}f^{G}_{IIIZ} \\ 
f^{G}_{XZII}f^{G}_{XZII} \\ 
f^{G}_{XIII}f^{G}_{XIII} \\ 
f^{G}_{XIIZ}f^{G}_{XIIZ} \\ 
\end{pmatrix} \) \\
\end{tabular}
\end{ruledtabular}
\end{table*}

\begin{table*}
\caption{\label{tab:4qo-gbg} Four-qubit open chains (green-blue-greem}
\begin{ruledtabular}
\begin{tabular}{cccccc}
        \multirow{4}{*}{\includegraphics[width=1.5cm]{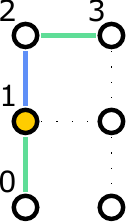}}& $F^B_1$ & $F_2^B$  & $\epsilon^B$ & $\vec{\Sigma}^B$  & $\vec{F}^B$ \\[.15cm]
        & $f^{B}_{IXII}$ & $f^{B}_{IXIX}f^{B}_{ZXIX}$  & $+0.50$ & \(\displaystyle \begin{pmatrix}
-0.25 \\ 
 +0.25 \\ 
 -0.50 \\ 
\end{pmatrix} \)  & \(\displaystyle \begin{pmatrix}
f^{B}_{ZXII}f^{B}_{ZXZI} \\ 
f^{B}_{IXZI}f^{B}_{IXII} \\ 
f^{B}_{IIIX}f^{B}_{IIIX} \\ 
\end{pmatrix} \) \\[.15cm]
\cline{2-6} \\[-.25cm]
& $F^G_1$ & $F_2^G$  & $\epsilon^G$ & $\vec{\Sigma}^G$  & $\vec{F}^G$ \\[.15cm]
        & $f^{G}_{IIXI}$ & $f^{G}_{XZXZ}f^{G}_{XZXI}$  & $-0.50$ & \(\displaystyle \begin{pmatrix}
-0.50 \\ 
 +0.50 \\ 
 +1.00 \\ 
\end{pmatrix} \)  & \(\displaystyle \begin{pmatrix}
f^{G}_{IIZI}f^{G}_{IIZI} \\ 
f^{G}_{IXZI}f^{G}_{ZXZI} \\ 
f^{G}_{IIZX}f^{G}_{IIIX} \\ 
\end{pmatrix} \) \\
\cline{1-6} \\

        \multirow{4}{*}{\includegraphics[width=1.5cm]{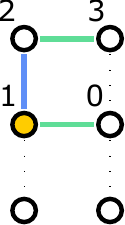}}& $F^B_1$ & $F_2^B$  & $\epsilon^B$ & $\vec{\Sigma}^B$  & $\vec{F}^B$ \\[.15cm]
        & $f^{B}_{IXII}$ & $f^{B}_{IXIX}f^{B}_{ZXIX}$  & $+0.50$ & \( \displaystyle 
        \begin{pmatrix}
+0.25 \\ 
 -0.25 \\ 
 -0.25 \\ 
 +0.25 \\ 
 -0.25 \\ 
\end{pmatrix}
         \)  & \(\displaystyle \begin{pmatrix}
f^{B}_{ZIII}f^{B}_{ZIII} \\ 
f^{B}_{IIIX}f^{B}_{IIIX} \\ 
f^{B}_{ZXII}f^{B}_{ZXZI} \\ 
f^{B}_{IXZI}f^{B}_{IXII} \\ 
f^{B}_{ZIIX}f^{B}_{ZIIX} \\ 
\end{pmatrix} \) \\[.15cm]
\cline{2-6} \\[-.25cm]
& $F^G_1$ & $F_2^G$  & $\epsilon^G$ & $\vec{\Sigma}^G$  & $\vec{F}^G$ \\[.15cm]
        & $f^{G}_{IIXI}$ & $f^{G}_{XZXZ}f^{G}_{XZXI}$  & $-0.50$ & \(\displaystyle \begin{pmatrix}
-0.25 \\ 
 +0.25 \\ 
 +0.50 \\ 
 -0.50 \\ 
 +0.75 \\ 
\end{pmatrix} \)  & \(\displaystyle \begin{pmatrix}

f^{G}_{ZIII}f^{G}_{ZIII} \\ 
f^{G}_{ZIIX}f^{G}_{ZIZX} \\ 
f^{G}_{IXZI}f^{G}_{ZXZI} \\ 
f^{G}_{IIZI}f^{G}_{IIZI} \\ 
f^{G}_{IIZX}f^{G}_{IIIX} \\ 
\end{pmatrix} \) \\
\end{tabular}
\end{ruledtabular}
\end{table*}

\begin{table*}
\caption{\label{tab:5qo} Five-qubit open chains}
\begin{ruledtabular}
\begin{tabular}{cccccc}
        \multirow{4}{*}{\includegraphics[width=1.5cm]{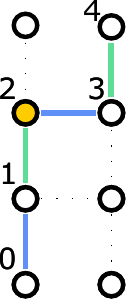}}& $F^B_1$ & $F_2^B$  & $\epsilon^B$ & $\vec{\Sigma}^B$  & $\vec{F}^B$ \\[.15cm]
        & $f^{B}_{IXIII}$ & $f^{B}_{IXIXI}f^{B}_{ZXIXZ}$  & $+0.50$ & \(\displaystyle \begin{pmatrix}
-0.25 \\ 
 +0.25 \\ 
 -0.50 \\ 
\end{pmatrix} \)  & \(\displaystyle \begin{pmatrix}
f^{B}_{ZXIII}f^{B}_{ZXZII} \\ 
f^{B}_{IXIII}f^{B}_{IXZII} \\ 
f^{B}_{IIIXI}f^{B}_{IIIXZ} \\ 
\end{pmatrix} \) \\[.15cm]
\cline{2-6} \\[-.25cm]
& $F^G_1$ & $F_2^G$  & $\epsilon^G$ & $\vec{\Sigma}^G$  & $\vec{F}^G$ \\[.15cm]
        & $f^{G}_{IIIXI}$ & $f^{G}_{IXZXZ}f^{G}_{ZXZXI}$  & $-0.50$ & \(\displaystyle \begin{pmatrix}
-0.50 \\ 
 +0.75 \\ 
 +0.25 \\ 
 +0.50 \\ 
\end{pmatrix} \)  & \(\displaystyle \begin{pmatrix}
f^{G}_{IIZII}f^{G}_{IIZII} \\ 
f^{G}_{IIZXI}f^{G}_{IIIXI} \\ 
f^{G}_{IIIXZ}f^{G}_{IIZXZ} \\ 
f^{G}_{IXZII}f^{G}_{ZXZII} \\ 
\end{pmatrix} \) \\
\cline{1-6} \\

        \multirow{4}{*}{\includegraphics[width=1.5cm]{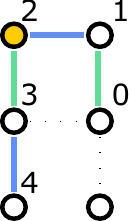}}& $F^B_1$ & $F_2^B$  & $\epsilon^B$ & $\vec{\Sigma}^B$  & $\vec{F}^B$ \\[.15cm]
        & $f^{B}_{IXIII}$ & $f^{B}_{IXIXI}f^{B}_{ZXIXZ}$  & $+0.50$ & \(\displaystyle 
        \begin{pmatrix}
+0.25 \\ 
 -0.25 \\ 
 -0.25 \\ 
 +0.25 \\ 
 -0.25 \\ 
\end{pmatrix}
         \)  & \(\displaystyle \begin{pmatrix}
f^{B}_{IXZII}f^{B}_{IXIII} \\ 
f^{B}_{ZIIXI}f^{B}_{ZIIXZ} \\ 
f^{B}_{ZXIII}f^{B}_{ZXZII} \\ 
f^{B}_{ZIIII}f^{B}_{ZIIII} \\ 
f^{B}_{IIIXZ}f^{B}_{IIIXI} \\ 
\end{pmatrix} \) \\[.15cm]
\cline{2-6} \\[-.25cm]
& $F^G_1$ & $F_2^G$  & $\epsilon^G$ & $\vec{\Sigma}^G$  & $\vec{F}^G$ \\[.15cm]
        & $f^{G}_{IIIXI}$ & $f^{G}_{IXZXZ}f^{G}_{ZXZXI}$  & $-0.50$ & \(\displaystyle \begin{pmatrix}
+0.25 \\ 
+ 0.25 \\ 
 -0.50 \\ 
 +0.50 \\ 
 -0.25 \\ 
 +0.50 \\ 
\end{pmatrix} \)  & \(\displaystyle \begin{pmatrix}
f^{G}_{IIIXZ}f^{G}_{IIZXZ} \\ 
f^{G}_{ZIIXI}f^{G}_{ZIZXI} \\ 
f^{G}_{IIZII}f^{G}_{IIZII} \\ 
f^{G}_{IXZII}f^{G}_{ZXZII} \\ 
f^{G}_{ZIIII}f^{G}_{ZIIII} \\ 
f^{G}_{IIIXI}f^{G}_{IIZXI} \\ 
\end{pmatrix} \) \\

\\
\cline{1-6} \\

        \multirow{4}{*}{\includegraphics[width=1.5cm]{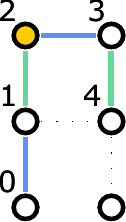}}& $F^B_1$ & $F_2^B$  & $\epsilon^B$ & $\vec{\Sigma}^B$  & $\vec{F}^B$ \\[.15cm]
        & $f^{B}_{IXIII}$ & $f^{B}_{IXIXI}f^{B}_{ZXIXZ}$  & $+0.50$ & \(\displaystyle 
        \begin{pmatrix}
+0.50 \\ 
 -0.50 \\ 
 -0.25 \\ 
 -0.25 \\ 
 +0.25 \\ 
\end{pmatrix}
         \)  & \(\displaystyle \begin{pmatrix}
f^{B}_{IXZII}f^{B}_{IXIII} \\ 
f^{B}_{IIIXI}f^{B}_{IIIXZ} \\ 
f^{B}_{IXIIZ}f^{B}_{IXZIZ} \\ 
f^{B}_{ZXIII}f^{B}_{ZXZII} \\ 
f^{B}_{IIIIZ}f^{B}_{IIIIZ} \\ 
\end{pmatrix} \) \\[.15cm]
\cline{2-6} \\[-.25cm]
& $F^G_1$ & $F_2^G$  & $\epsilon^G$ & $\vec{\Sigma}^G$  & $\vec{F}^G$ \\[.15cm]
        & $f^{G}_{IIIXI}$ & $f^{G}_{IXZXZ}f^{G}_{ZXZXI}$  & $-0.50$ & \(\displaystyle \begin{pmatrix}
+0.25 \\ 
+ 0.50 \\ 
 -0.25 \\ 
 +0.25 \\ 
 +0.75 \\ 
 -0.25 \\ 
 -0.50 \\ 
\end{pmatrix} \)  & \(\displaystyle \begin{pmatrix}
f^{G}_{IXIIZ}f^{G}_{ZXIIZ} \\ 
f^{G}_{IXZII}f^{G}_{ZXZII} \\ 
f^{G}_{ZXIII}f^{G}_{IXIII} \\ 
f^{G}_{IIIXZ}f^{G}_{IIZXZ} \\ 
f^{G}_{IIIXI}f^{G}_{IIZXI} \\ 
f^{G}_{IIIIZ}f^{G}_{IIIIZ} \\ 
f^{G}_{IIZII}f^{G}_{IIZII} \\ 
\end{pmatrix} \) \\

\end{tabular}
\end{ruledtabular}
\end{table*}

\begin{table*}
\caption{\label{tab:4qc} Four-qubit closed chain}
\begin{ruledtabular}
\begin{tabular}{cccccc}
        \multirow{4}{*}{\includegraphics[width=1.5cm]{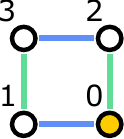}}& $F^B_1$ & $F_2^B$  & $\epsilon^B$ & $\vec{\Sigma}^B$  & $\vec{F}^B$ \\[.15cm]
        & $f^{B}_{IXII}$ & $f^{B}_{IXXX}f^{B}_{IYYX}$  & $+0.50$ & \(\displaystyle \begin{pmatrix}
+0.25 \\ 
 -0.50 \\ 
 -0.25 \\ 
 -0.25 \\ 
 -0.25 \\ 
 -0.25 \\ 
 -0.25 \\ 
 +0.50 \\ 
 +0.25 \\ 
 +0.25 \\ 
 +0.50 \\ 
 -0.25 \\ 
\end{pmatrix} \)  & \(\displaystyle \begin{pmatrix}
f^{B}_{IIXY}f^{B}_{IIXY} \\ 
f^{B}_{IIXY}f^{B}_{IIYX} \\ 
f^{B}_{IXIX}f^{B}_{ZXZX} \\ 
f^{B}_{ZIII}f^{B}_{ZIII} \\ 
f^{B}_{IIXX}f^{B}_{IIXX} \\ 
f^{B}_{IYIX}f^{B}_{ZYZX} \\ 
f^{B}_{IIZI}f^{B}_{IIZI} \\ 
f^{B}_{IIZX}f^{B}_{IIIX} \\ 
f^{B}_{ZIZI}f^{B}_{ZIZI} \\ 
f^{B}_{IXII}f^{B}_{ZXII} \\ 
f^{B}_{IXII}f^{B}_{ZYII} \\ 
f^{B}_{ZYII}f^{B}_{IYII} \\ 
\end{pmatrix} \) \\[.15cm]
\cline{2-6} \\[-.25cm]
& $F^G_1$ & $F_2^G$  & $\epsilon^G$ & $\vec{\Sigma}^G$  & $\vec{F}^G$ \\[.15cm]

        & $f^{G}_{IIXI}$ & $f^{G}_{ZXYY}f^{G}_{ZYXY}$  & $-0.50$ & \(\displaystyle \begin{pmatrix}
-0.25 \\ 
 +0.25 \\ 
 +0.25 \\ 
 -0.25 \\ 
 +0.25 \\ 
 +0.25 \\ 
 +0.50 \\ 
 -0.25 \\ 
 -0.25 \\ 
 +0.25 \\ 
 +0.50 \\ 
 -0.25 \\ 
 +0.25 \\ 
 -0.50 \\ 
 -0.25 \\ 
 +0.25 \\ 
\end{pmatrix} \)  & \(\displaystyle \begin{pmatrix}
f^{G}_{ZZII}f^{G}_{ZZII} \\ 
f^{G}_{IXIY}f^{G}_{IXIY} \\ 
f^{G}_{IIXY}f^{G}_{ZZXY} \\ 
f^{G}_{IIYI}f^{G}_{ZIYI} \\ 
f^{G}_{ZXII}f^{G}_{ZXIZ} \\ 
f^{G}_{ZYII}f^{G}_{ZYIZ} \\ 
f^{G}_{IXIX}f^{G}_{IYIY} \\ 
f^{G}_{IXIZ}f^{G}_{IXII} \\ 
f^{G}_{ZIII}f^{G}_{ZIII} \\ 
f^{G}_{IZII}f^{G}_{IZII} \\ 
f^{G}_{IIXI}f^{G}_{ZIYI} \\ 
f^{G}_{IXIX}f^{G}_{IXIX} \\ 
f^{G}_{ZIXI}f^{G}_{IIXI} \\ 
f^{G}_{IIIY}f^{G}_{IZIY} \\ 
f^{G}_{IYII}f^{G}_{IYIZ} \\ 
f^{G}_{IIYY}f^{G}_{ZZYY} \\ 
\end{pmatrix} \) \\
\cline{1-6} \\

\end{tabular}
\end{ruledtabular}
\end{table*}